\documentclass[12pt,oneside,english]{article}

\usepackage{amsmath}
 \usepackage{graphicx}
\usepackage[T1]{fontenc}
\usepackage[latin1]{inputenc}
\usepackage{babel}
\usepackage{setspace}
\usepackage{rotating}

\makeatletter

\begin{document}

\centerline{ \bf Double-slit interference pattern from single-slit screen and
its gravitational analogues}

\bigskip \bigskip

\centerline{\large  D.  Bar} 

\bigskip

\begin{abstract}
{\it The double slit experiment (DSE) is known as an
important cornerstone in the foundations of physical theories such as 
Quantum Mechanics and Special Relativity. A large number of different variants
of it were designed and performed over the years. We perform and discuss  
here a new verion
 with the somewhat unexpected results of obtaining  interference
pattern from single-slit screen.  We show using either 
  the  Brill's version 
 of the canonical
 formulation of general relativity  or  the linearized version of it that one
 may find corresponding and analogous situations in the framework of 
 general relativity.    }

\end{abstract}

\bigskip 

\noindent \underline{ Pacs numbers}: $ \ \ $  42.25.Hz, 04.20.Gz, 04.30.-w, 04.20.Fy, 04.30.Nk 
 
     \bigskip

     \bigskip 
     
     \markright{INTRODUCTION} 
     
\protect \section{\bf INTRODUCTION}

There is no physical experiment which plays such an important historical  
 role in
the establishment and development of fundamental physical theories \cite{shamos}
such as the double slit experiment (DSE). It was first represented in 1803 by Young
\cite{young} as a desicive proof of the wave nature of light. Later, its
interferometric character serves,  through the
Michelson-Morley experiment \cite{michelson,bergmann}, as the key trigger 
 for the enunciation of 
 the special
relativity and  the constancy of light velocity \cite{bergmann}. Still later,
experimenting with some variants of the DSE (see the following paragraphs) has
aroused  fundamental problems and paradoxes which lead to a new physical
understanding embodied in the laws of Quantum Mechanis (\cite{merzbacher}, 
see, especially, chapter
1 in \cite{schiff}).  \par   
As known,   the interference pattern resulting  from the DSE 
\cite{schiff,jenkins}  appears even if the intensity of the passing light is
decreased until an average of only one photon is in transit between 
source  and double-slit 
screen \cite{schiff}. From this one may conclude \cite{schiff} that a single
photon is capable of interfering with itself. Moreover, although each of the 
passing photons can go
only through one of the slits the interference pattern appears 
only when both
slits are open \cite{schiff,merzbacher,wheeler,zajonc}. In other words, as emphasized in
\cite{schiff,wheeler,zajonc}, if the experiment is also designed,  by 
 either closing one slit or   adding photons detectors at the slits 
(see Figure
3 in \cite{schiff}), to  supply 
the additional information of the exact slit through which each photon passes 
 then the interference pattern disappears 
\cite{schiff,merzbacher,wheeler,zajonc}. This demonstrates the large influence
of the observation itself upon the obtained results (see, for example,
\cite{wheeler}, for this influence in Quantum Mechanics and see
\cite{finkelstein} for this influence in relativity).  \par 
In order to isolate the 
real factors which
determine the presence or absence of the interference pattern in the DSE  we
 summarize  here some versions  of it \cite{wheeler,zajonc,yoon} with particle
detectors at the slits.  \par 
\noindent \underline{Experiments}:
The particle detectors  are; \par 
\noindent  $(a)$ Turned-on and the data about the exact routes passed by the
photons recorded and {\it used during the experiment}. \par
\noindent  $(b)$ Turned off and, therefore, no data were recorded and used during the 
 experiment.  \par
\noindent $(c_1)$ Turned on but the
observer does not bother to record the count at the slits. \par
\noindent $(c_2)$ Turned on and also recording the count at the
slits but this information is  thrown and {\it not used during the experiment}. \par
\noindent $(d)$  Turned-on and also recording the  supplied information  
 but  it is 
 mixed with other unrelated data so that   
 the observer  prepares 
 some program which  analyzes the combined information  in either one
 of the two following ways;  \par  
\noindent   $(d_1)$  The unrelated data are removed in which case one
 remains with the real data from the detectors.  \par 
 \noindent  $(d_2)$ 
  Keeping the whole
 mixed-up information  so the true data from the detectors are not
 available.  
\par
\noindent \underline{Results}: For final analysis of $(b), \ (c_1), \ (c_2), \
(d_2)$ 
an interference pattern
 appears and for $(a), \ (d_1)$  no
 such pattern  is seen.  We note that for $(a)$, $(d_1)$ the optical pattern
 shown on the screen is separated into two  patterns
 \cite{wheeler,zajonc,yoon} each of them is of the single-slit experiment (SSE)
 kind and 
 corresponds  to the slit which is in line with it. That is, there exists no
 interference of any photon from one slit with any other photon from the other
 slit so that each of the two patterns is formed from the diffraction of the
 particles which pass through the slit  in line with it   as schematically shown
 in Figure 3.   
  \par 
One may realize from $(b)$, $(c_1)$, $(c_2)$ and   $(d_2)$  that even 
 the determination and recording of the exact routes of 
the particles through the double-slit screen are not enough 
 for cancelling   the interference  pattern  
 if, as mentioned, 
 these data from the detectors are not   included in the experiment 
 itself \cite{schiff}.    
  It can also be seen    from the results of $(d_1)$ and $(d_2)$ that  the 
  inclusion of
  these data in 
  the
  DSE  may be performed   
    even years   
   after completing  the  passage through the slits (of course, before obtaining
   the optical pattern on the photosensitive screen)  
   which
 constitutes the known delayed choice experiment \cite{wheeler,yoon}. That is,
 the  important factor which determines the form of the obtained optical
 pattern is the use (or not),  during the DSE,  of the information about 
 the routes of
 the photons through the slits.   Moreover, as seen  in the optical literature
  \cite{jenkins},  this is valid for any multiple-slit screen. That is, any such
  screen   may
  demonstrate either the interference pattern of the DSE kind (as  in Figure 1) 
  if the data about
  the routes through slite are {\it not used during the experiment} or a number
  of diffraction patterns each of them of the SSE  kind (see
  Figure 2) if these data
  are used. The number of these SSE's diffraction patterns equals the number of
  slits as seen, for example, in Figure 4 for the four-slit screen where the
  data about routes are used during the experiment. The only difference between
  the interference pattern of the DSE and that of 
  any other $n$-slit pattern  is that the larger is $n$ the thinner become  
  the fringes of this pattern \cite{jenkins}. \par  
  We show,  through actual experiments,   that  the noted conditions of 
  using or
  not using the data about the photons routes which, respectively,  entail
    diffraction or interference patterns   are 
 valid not
 only for the  
 double
 or any other multiple-slit screen but also  for 
  the  SSE. Note that for any $n$-slit screen, where $n \geq 2$, the 
  mentioned
  data
  about routes may sum,    
  for a large number
  of photons,   to a huge amount of information whereas   for 
  the  SSE these data  involve only one     
   single piece   which is that all the photons pass only
  through that slit.  Thus, we show  that if this 
  single data  is {\it not used  during the
  experiment} then the expected single-slit diffraction pattern does not appear.
  \par 
   It should be
  noted in this context that up to now the very nature of the employed 
  screen, if it
  is single, double or multiple-slit,    is always known and used during the
   experiment. But unlike the double and any other multiple-slit experiments  
    for
   which one may know and, therefore, use during the experiment the datum  
   about  the number of 
   slits without   using  the data about
   the routes of the photons through them  
   for  the SSE case  they, actually, lead to each other.
   This is because if one knows and, therefore, use during the experiment the
   datum that he is employing single-slit screen then he, automatically, also
   knows and use the datum that all photons pass through that slit and vice
   versa.  \par  
       Suppose, now, that the observer activates a single-slit screen without
       knowing it  and, therefore, the  information related to 
       the single route of all the photons  can not be used during 
       the experiment.            
          We show         
    that in this case, and under the   special conditions  described in 
     Section II,   
    the 
 resulting pattern is that of  interference as demonstrated in the appended
 pictures \cite{camera}. 
 That is, instead of obtaining the diffraction pattern,  shown in  
 Figure 2,  which
 is typical of SSE we have obtained the 
 interference pattern, shown in   Figure 1,  
  which is characteristic of $n$-slit screen ( $n \geq 2$).      
 That is,  one may conclude 
 for any $n$-slit screen, where  $n \geq  1$,  that if one uses  
  the data
 about the routes then one  obtains $n$ diffraction patterns  where $n$ is
  the number of slits
 as shown, for example, in Figure 2 for $n=1$ and in Figures 3-4 for $n=2$
 and $n=4$. 
  If, however, these data are not taken into account
    during the experiment then the obtained pattern for any $n$-slit screen,
    even for $n=1$,  
     is the
    interference  one shown in  Figure 1. In other words,
    for all  $n$-slit screens, where $n \geq 1$,  changing the situation  
    from using during the experiment the data about the photon routes to not
    using them amounts to changing these routes from being densely and
    continuously arrayed in the
    forward directions ($m=0$) to being fringed and striped. 
     \par
      We show in this work 
    that this formation   
    of   periodic fringes and stripes   
    may also be found as geodesics changes in at least two theoretical branches 
     of the 
    general relativity theory. As known, 
     using the equivalence principle \cite{bergmann,mtw,hartle} one may
    discuss any physical event as either occuring in a flat spacetime with
    physical interactions or resulting from a curved spacetime
    {\cite{bergmann,mtw,hartle} with no such
    interactions. Thus, the mentioned change of the  photon routes, from 
    the continuous diffraction type (in the neighbourhood of the order $m=0$) 
     to
    the fringed interference one,  may be, theoretically, paralleled to    
    corresponding  situations in general relativity  
     \cite{bergmann,mtw,hartle}.  \par 
     Note that one may argue  that the mentioned DSE results, 
     detaily described in Section II,
     should  be exclusively discussed in pure quantum mechanical terms without
     having to invoke any general relativity idea. We answer to this
      that  the 
     relativistic  discussion here is not suggested as some kind of 
     explanation or
     interpretation of this DSE. Our aim in this discussion is to point out that
     corresponding and parallel situations may also, as mentioned, be
     encountered in the framework of general relativity. That is, one may,
     theoretically,  find formations of fringed and nonfringed trapped surfaces
     which are related to the same kind of GW (either the Brill or plane GW's)
     as those found with the same kind of optical screen (see Figures 7-8). 
     \par 
    As known \cite{bergmann,mtw,hartle}  geodesics  changes
    are related in the general relativity to corresponding changes in the
    geometry of the surrounding spacetime
    which are, especially, tracked to the presence of gravitational 
    waves (GW) 
     \cite{mtw,hartle,thorne}. Moreover, if these GW are strong enough they 
     may entail 
    corresponding and lasting changes in the form of the relevant geodesics
    which stay  long after the generating GW disappear. 
    As mentioned, 
   the noted changes in the photon routes  
   result only from
  considering (or not)  during the experiment the data regarding these routes 
  and not
  from any other  force. Thus, a suitable parallel gravitational situation 
  is  related   more to   the pure source-free GW's  
  \cite{mtw,brill1,eppley,miyama}  than    to the 
  matter-sourced ones \cite{mtw,hartle,thorne}. Accordingly, we pay special
  attention  in the following  
   to these 
  source-free gravitational fields      
   which  
   constitute solutions to the Einstein vacuum field equations  
    \cite{bergmann,mtw,hartle}  and  propagate in vacuum as 
    pure source-free  
 GW's  \cite{mtw,brill1,eppley,gentle} 
  with no involvement of matter.  \par  As representatives of these 
  source-free
    radiation  we  consider the (1) Brill GW's
  \cite{brill1,eppley,gentle,alcubierre} and (2) the 
   plane 
  GW's in the linearized version of general relativity.  
  The later kind is chosen  because it is discussed in
  the almost flat metric which is similar to the flat metric of the
  mentioned optical experiments. Moreover, it has been 
  shown \cite{bar1,bar2} 
  that, 
  like the electromagnetic (EM) $n$-slit experiments,  the plane
  GW in the linearized version of general relativity have, under special
  conditions, properties which make it  
   capable of  interfering with other GW's. 
     Both of these GW  change, if they are strong
  enough, the surrounding spacetime and its topology  
  \cite{finkelstein1,sorkin} 
   which, naturally,
  entail also changes of geodesics. The new changed  
  spacetime is, theoretically,
  represented,  in the spacetime region traversed
  by these GW's,  by trapped surfaces
  \cite{mtw,eppley,brill2,beig,abrahams,alcubierre} 
  which   have the same intrinsic geometry as that
  of the generating GW. One may, especially, count two 
different kinds of these surfaces; (1) the  
singular trapped ones 
\cite{eppley,abrahams,alcubierre,tipler,yurtsever} and (2) the 
nonsingular ones
\cite{beig} which are 
related to the  regular and 
asymptotically
flat initial data \cite{mtw,eppley,brill1,adm} in vacuum. 
 Note that as \cite{bar1} 
    interference patterns and holographic images 
   result from interfering electromagnetic 
   waves 
   so  trapped surfaces result also from interfering GW's. \par
    
 In Section II we represent a detailed account of the  experiment  
  from which
 we obtain, under suitable conditions,  interference fringed  
  pattern from single-slit 
 screen.  
  In Section III we have shown that
 one may obtain similar fringed geodesics   
  by beginning from the
 Brill's metrics and use the terms and terminology  related to it. 
  We have, also,   calculated the corresponding 
 fringed trapped
 surface  which is formed from this Brill's pure radiation.  
 The trapped surface, without fringing, 
  resulting from the
 Brill GW's  have been calculated in 
 \cite{eppley} and
 were represented, for comparison and completness, in Appendix A.  
 In this
 Appendix we have also represent a short review of the ADM canonical 
 formulation \cite{adm,mtw} 
 of general relativity and the specific conditions which lead to 
 the Brill's
 source-free GW's. 
 In Section IV we 
 have calculated the  fringed trapped surface   
   obtained from  plane GW's by 
 using the approximate 
 linearized general relativity \cite{mtw}. The plane GW's   
 in the framework of this approximate theory   and the resulting 
 trapped surfaces, without fringing, 
  have been detaily calculated in \cite{bar1}
 and were represented, for comparison and completness, in Appendix B.  
   We discuss and summarize the main results  
 in a Concluding
 Remarks Section. \par

\begin{figure}
\begin{turn}{-90}
\includegraphics[width=5.9cm]{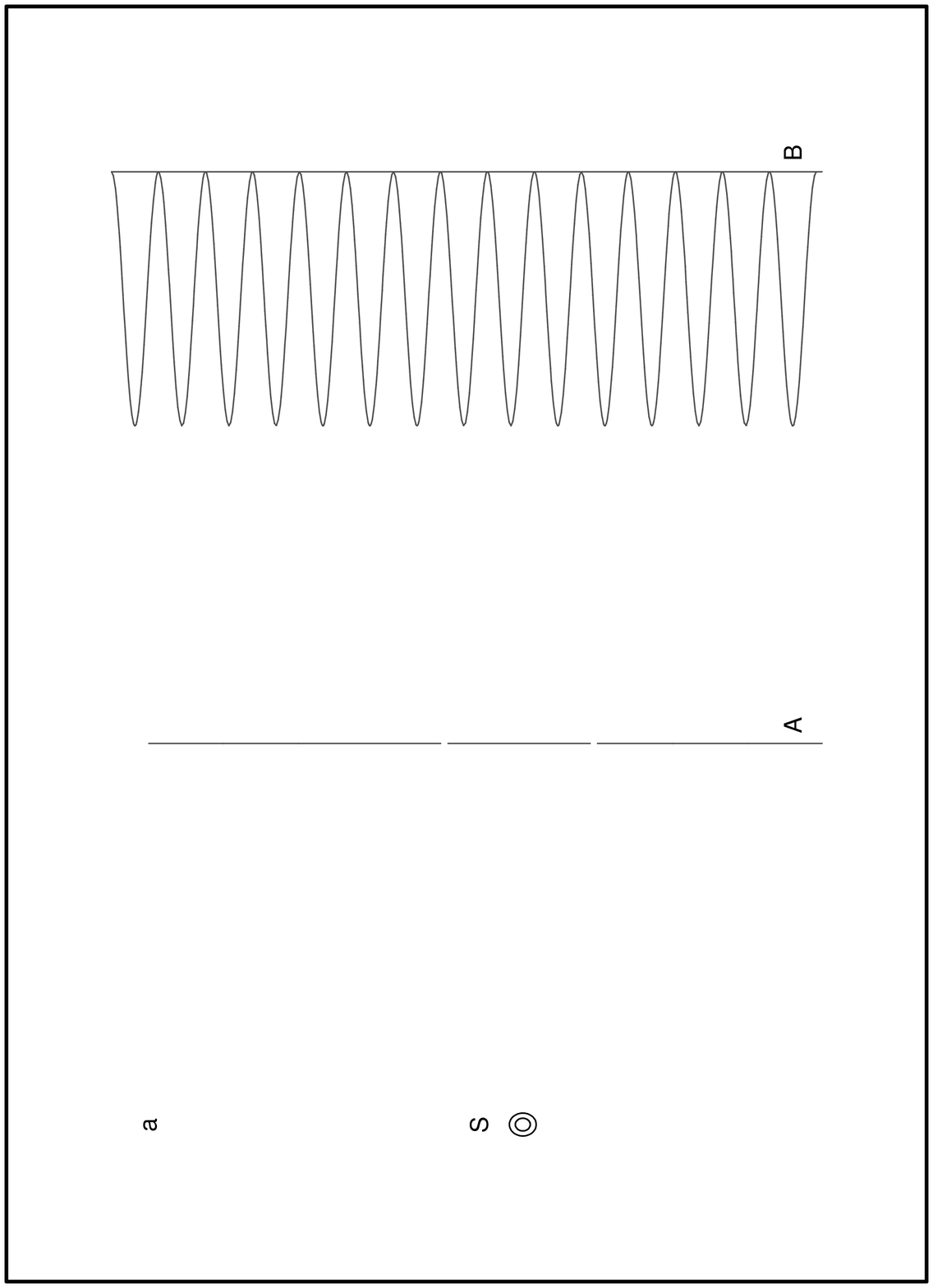}
\end{turn}
\begin{turn}{-90}
\includegraphics[width=5.9cm]{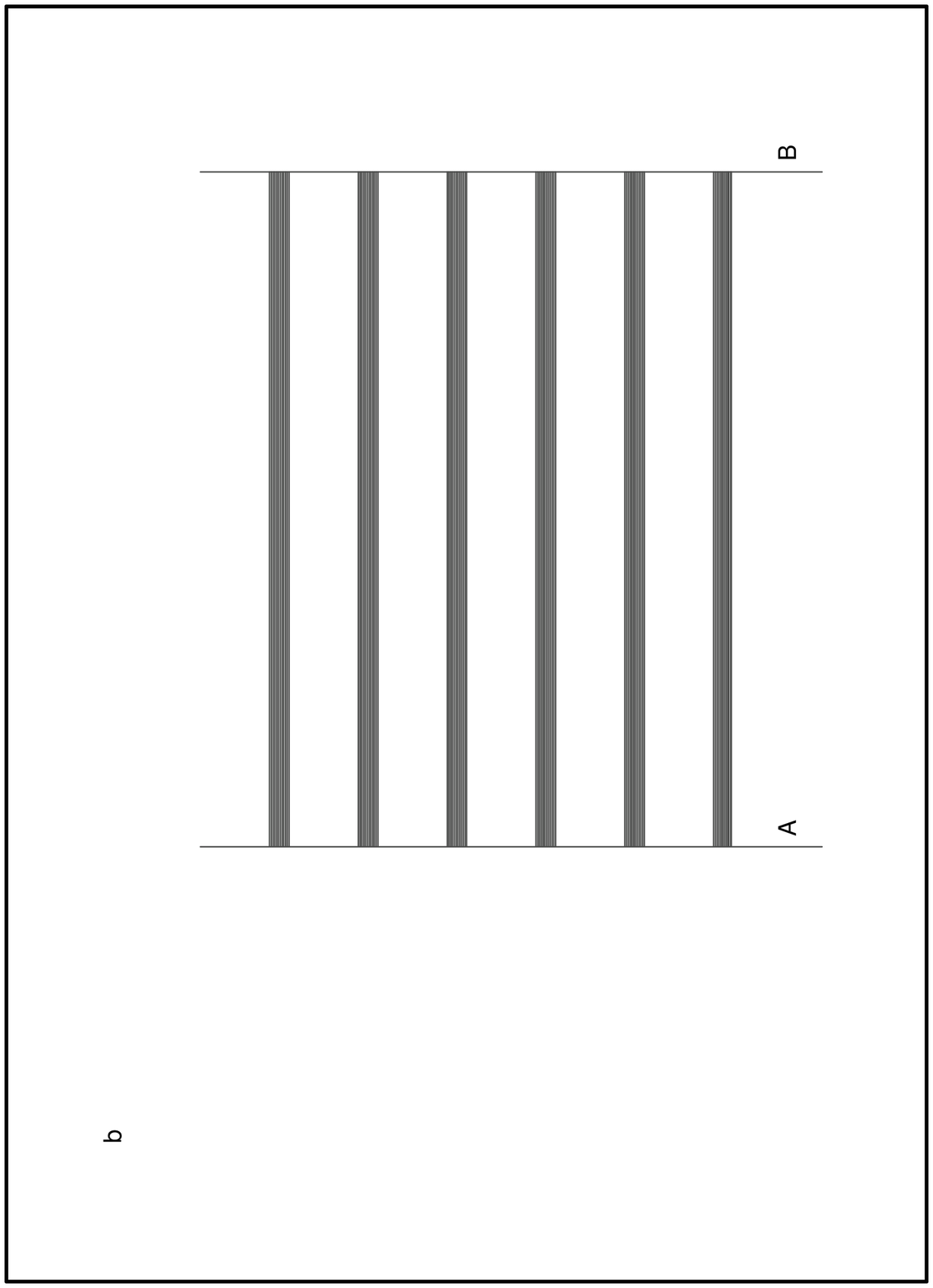}
\end{turn}

     \caption{A schematic representation of the DSE, including the light source
     $S$,  is shown at 
     the left Subfigure $a$ where the data about the routes through slits are
     not used during the experiment. 
      The interference pattern 
resulting from the optical path difference between the  rays 
from the  two slits at screen $A$  is shown  on the
 photosensitive screen $B$. At the right Subfigure $b$ one may see the 
 corresponding 
 routes traversed
 by the photons between the two screens.}
     \end{figure}

\begin{figure}
\begin{turn}{-90}
\includegraphics[width=5.9cm]{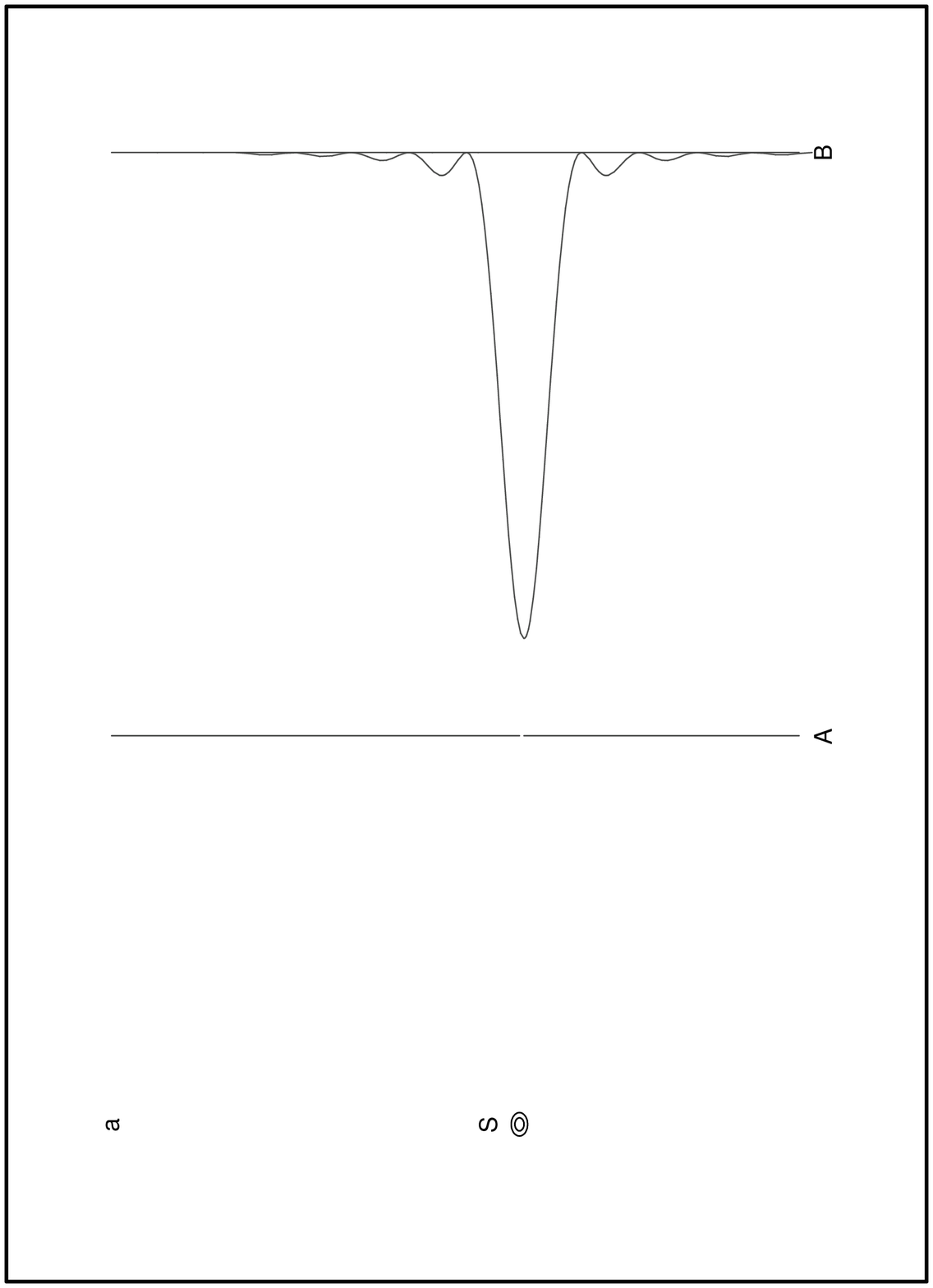}
\end{turn}
\begin{turn}{-90}
\includegraphics[width=5.9cm]{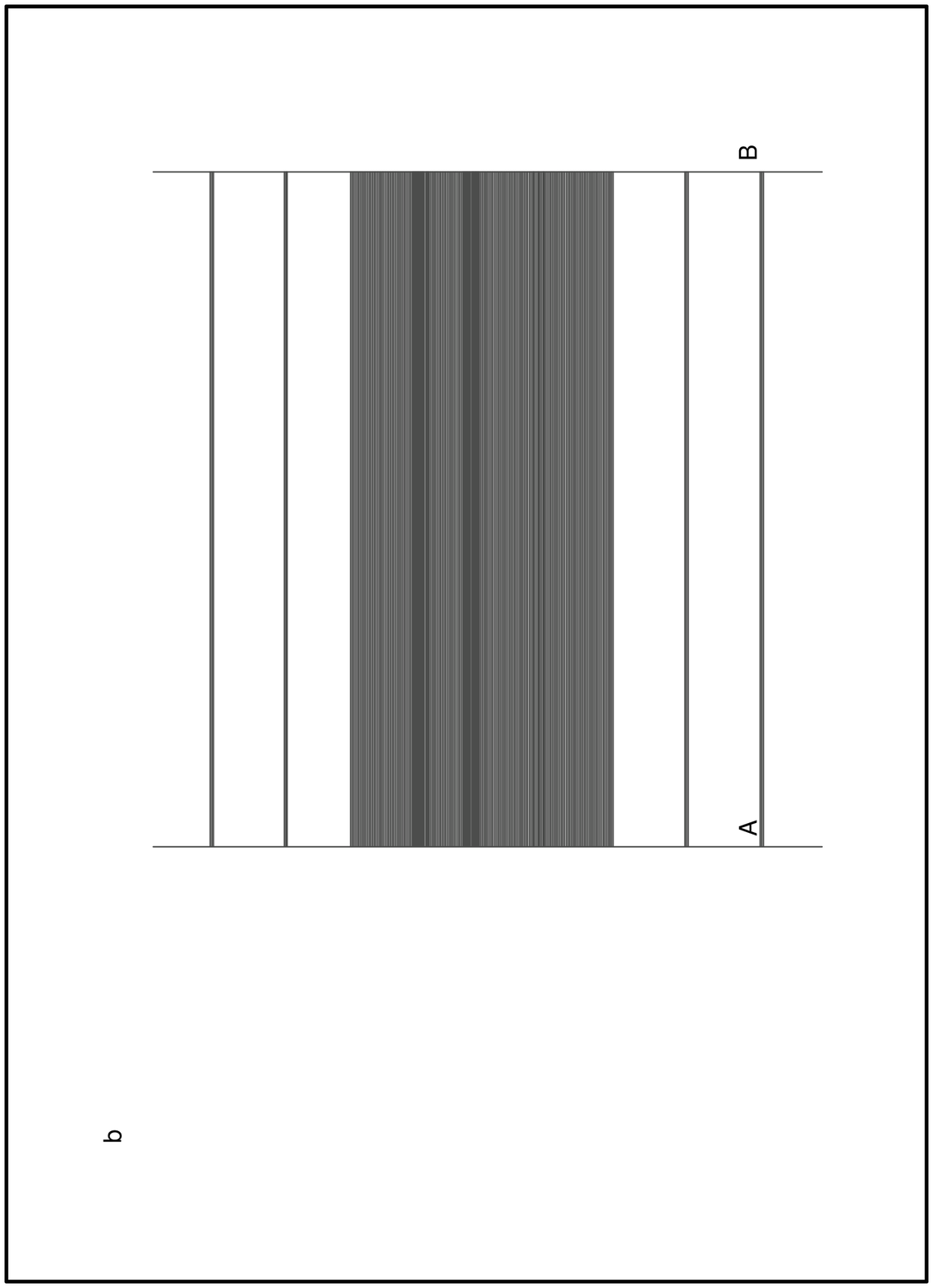}
\end{turn}

    \caption{A schematic representation of the SSE, including the light source
    $S$,   is shown at the 
    left Subfigure $a$ where the observer knows and, therefore, uses during the
    experiment the datum about the nature of screen. 
      The diffraction pattern 
resulting from the optical path difference between   rays 
from same slit at screen $A$  is shown  on the
 photosensitive screen $B$. At the right Subfigure $b$ one may see the 
 corresponding 
 routes traversed
 by the photons between the two screens where the most traversed route 
 is that in the forward direction   ($m=0$).}
     \end{figure}

\begin{figure}
\begin{turn}{-90}
\includegraphics[width=5.9cm]{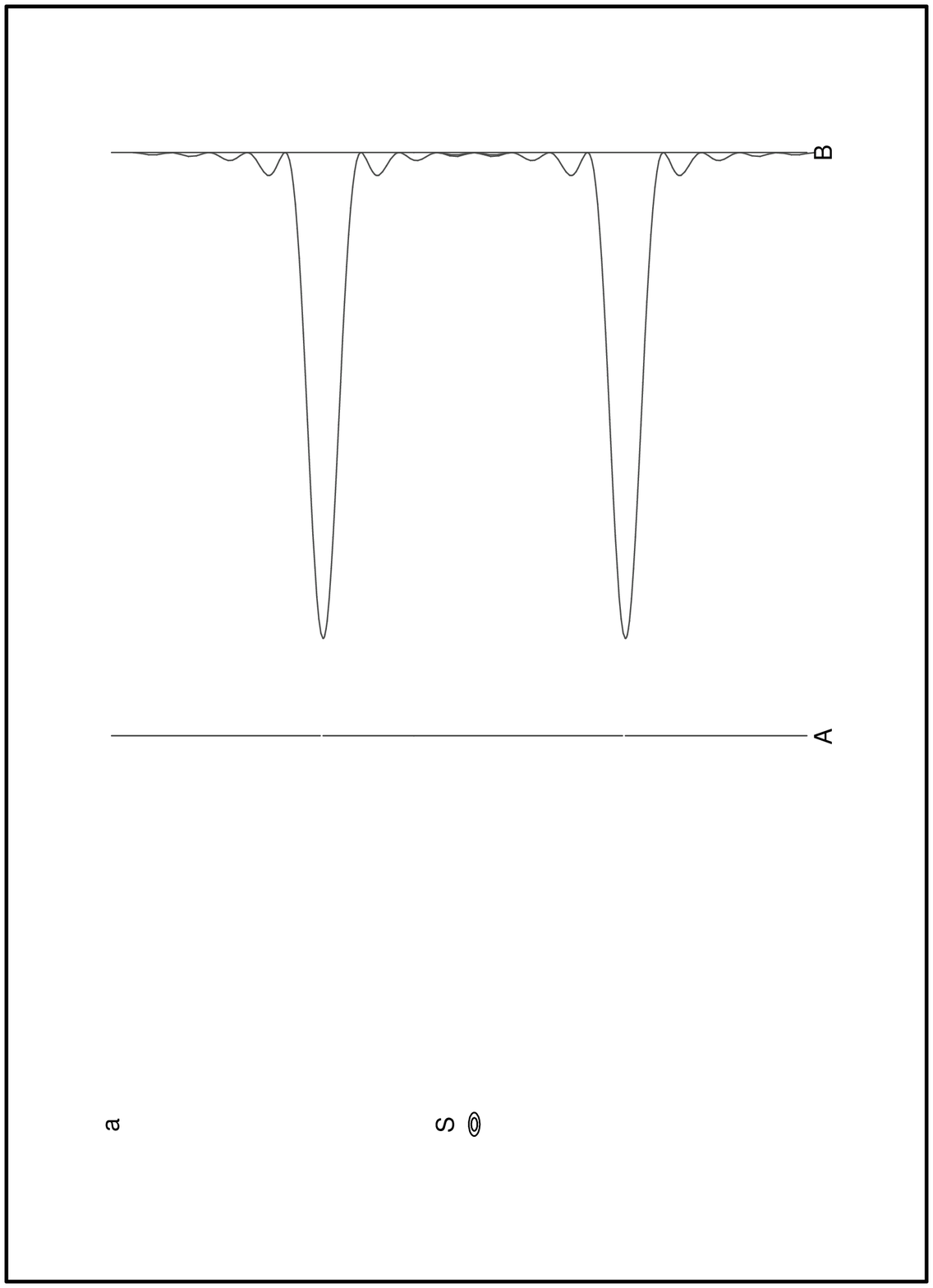}
\end{turn}
\begin{turn}{-90}
\includegraphics[width=5.9cm]{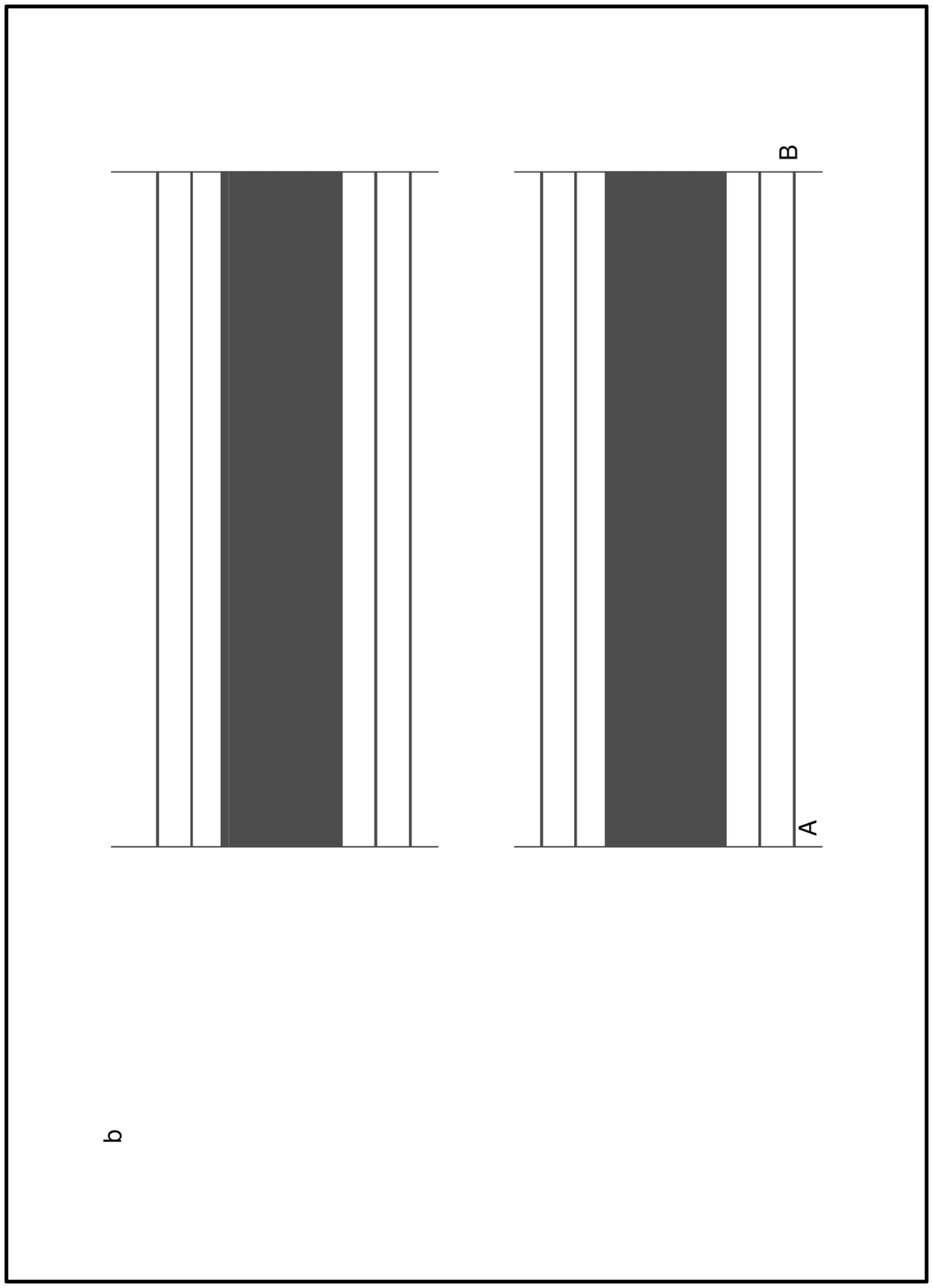}
\end{turn}

 \caption{At the left Subfigure $a$  we see the  optical pattern 
 obtained from the double-slit
 screen when the routes of the photons through the slits are recorded 
 and taken
 into accound during the experiment. One have here no interference 
 between rays from the two slits but
 two separate diffraction patterns  each composed  by the 
 photons passed through the slit in line with it. 
   At the right Subfigure  $b$ we see the form of the routes between the
  screens where most photon pass in the two forward directions.  
   }
  \end{figure}

\begin{figure}
\begin{turn}{-90}
\includegraphics[width=5.9cm]{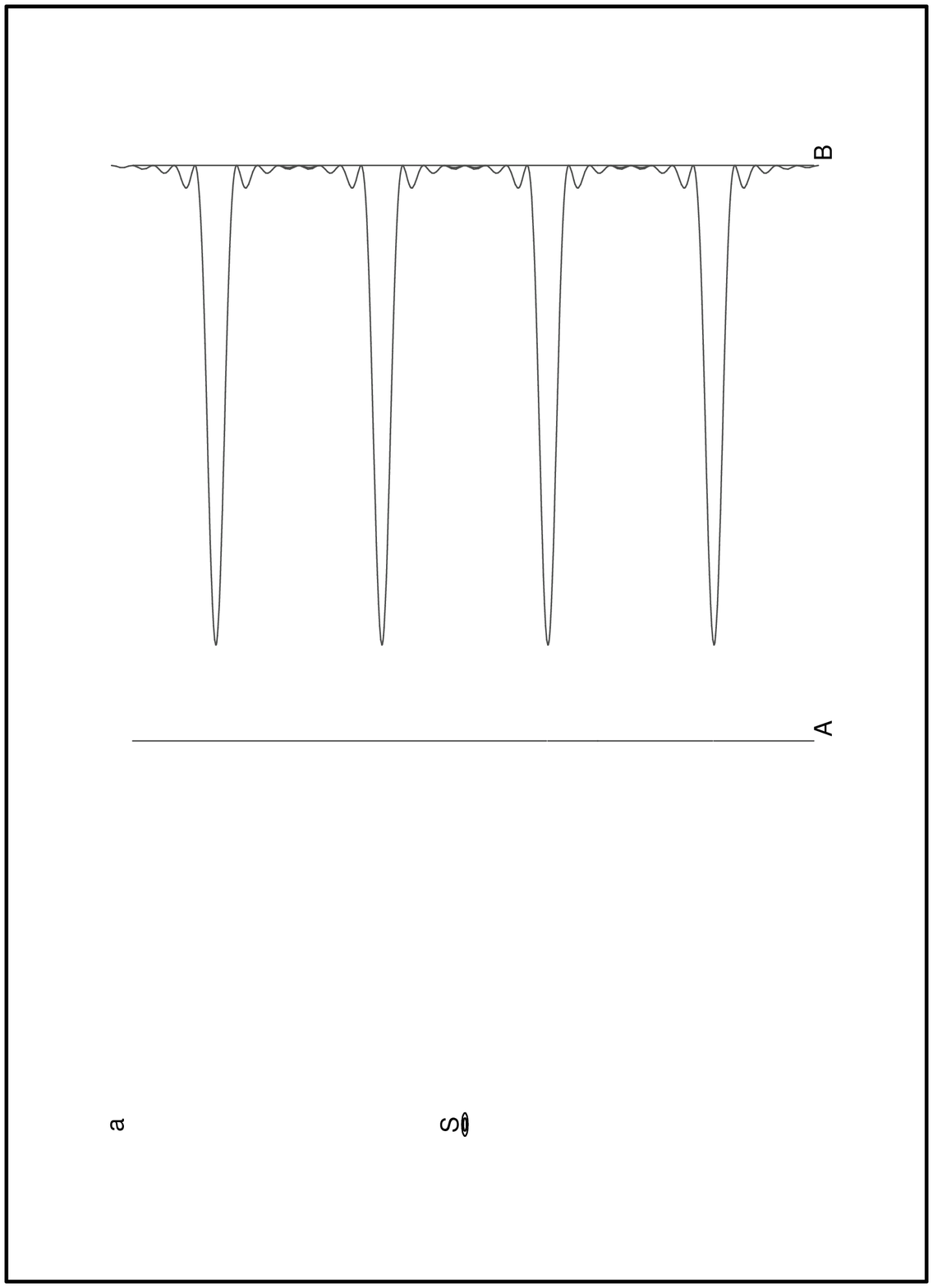}
\end{turn}
\begin{turn}{-90}
\includegraphics[width=5.9cm]{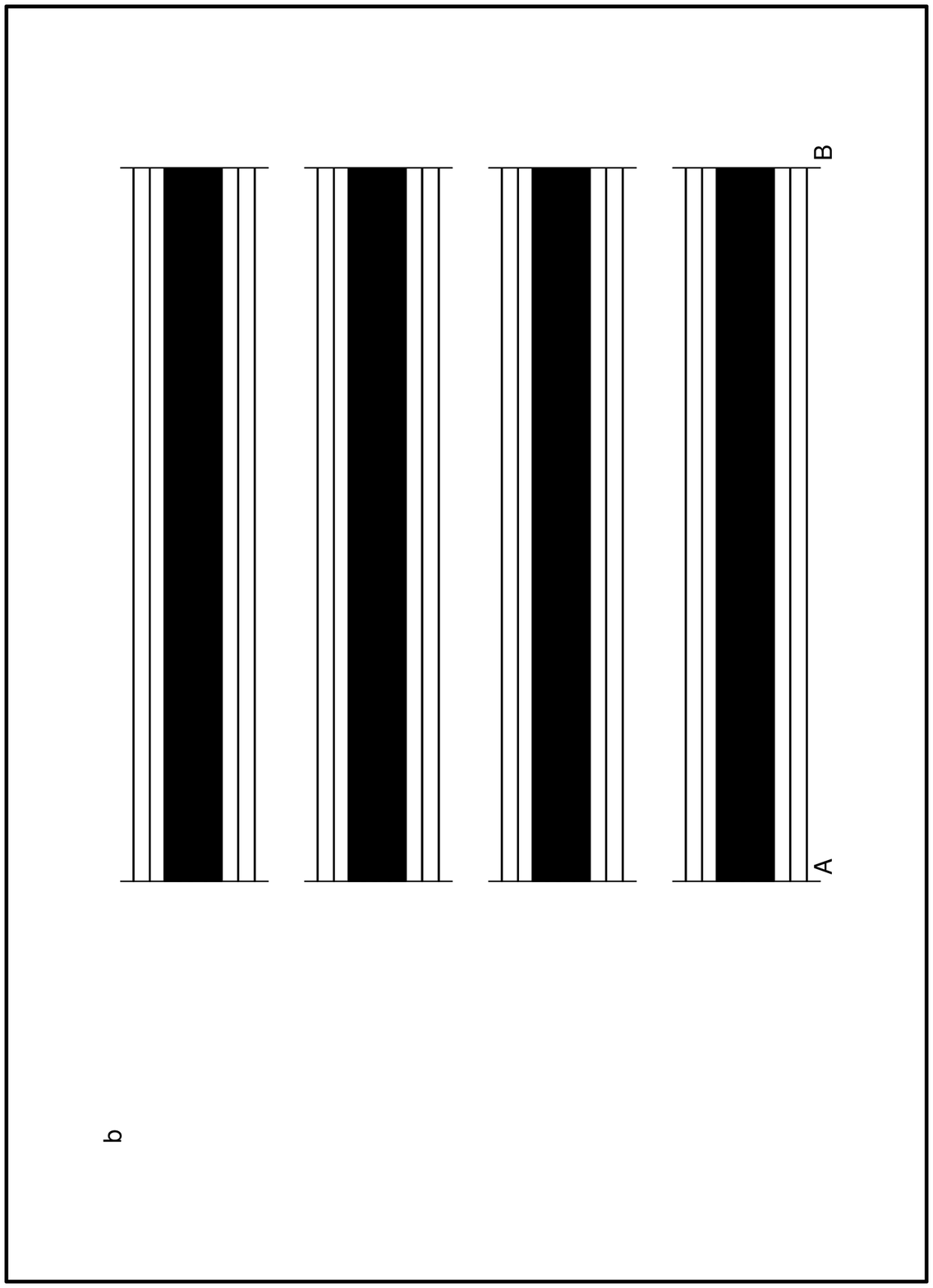}
\end{turn}

 \caption{At the left Subfigure $a$  we see the  optical pattern 
 obtained from the four-slit
 screen when the routes of the photons through the slits are recorded 
 and taken
 into accound during the experiment. One have here no interference 
 between rays from any two slits but
 four separate diffraction patterns  each composed  by the 
 photons passed through the slit in line with it. 
   At the right Subfigure $b$  we see the form of the routes between the
  screens where most photon pass in the four  forward directions.  
   }
  \end{figure}

 \markright{OBTAINIG INTERFERENCE PATTERN FROM SSE}  
  
\protect \section{Obtaining interference pattern from SSE}
The experimental set-up for the variant of the DSE discussed here includes 
a laser pointer as light source and 30 mirrors. The laser pointer acts as a
strong monochromatic red light source with wavelength in the 650-680 $nm$ range
and output of less than 1 $mW$. The mirrors were prefered to serve as double and
single slit screens because it is easy to inscribe on their coated sides very 
narrow
slits (scratchs).  Thus, using double edge razor blades
 two narrow slits of about $0.3 \ mm$ wide were cut in the coated side of 
 each
of them where the distances between the slits  varied in the range of $1-4 
 \
mm$.
Twenty nine   (29)  mirrors  were prepared to serve as real double slits
screens 
and in the remaining  mirror  
one  slit was real and the second was spurious and superficially  cut 
so that   it was still opaque. This is done so that when the observer
 looks at this mirror  from some distance he  
 would not differentiate between it and the other 
real  two-slit ones. \par
  In Figure 5 one may   see a  photograph of the actual arrangement of the
   experiment. This figure  as well as Figures 6-8 are real pictures 
   photographed
   by a digital camera \cite{camera}. In Figure 5 one may see at the front the
   laser  pointer mounted upon a white rectangular box. As seen, this laser
   pointer is hold by two binder clips which serve the twofold purpose of
   conveniently directing  it towards the back of the mirror-screen and also of
    pressing its operating button so as to activate it. At the
    back of the Figure one may see a second rectangular box upon which 
     one    mirror (of the available 30)  is  mounted  with the help of 
     a  second pair of
    binder clips. One may also see at the back
    of the mirror the real or spurious double slit 
     and the red laser ray from the pressed laser pointer. 
    Moreover, with a larger
    resolution one may even discern the red optical pattern at the white wall
    behind the mirror. At the right one may see inside another 
    rectangular box  the operated mirrors. In Figure 6 one may see a 
    typical picture of the optical
    interference pattern obtained from one of the 29 real  double-slit mirrors
    where the nature of this mirror is known and used during the experiment but
    not the data about the  routes of the photons through the slits. 
    In Figure 7
    we show two typical photographs of the optical diffraction pattern obtained
    from the faked double-slit mirror when its true nature, that it is
    single-slit, is used during the experiment. The left picture was
    photographed by   taking a longer distance between the mirror and the white
    wall compared to the distance used for the right picture. \par 
       Using random number generator the observer begins the experiment by 
randomely 
 picking one mirror from the
available 30 without knowing if he choose  one of 
the real double-slits or the
faked one where the probability  to choose  the  former is $\frac{29}{30}=0.97$  
and that for   picking the latter is 
 $\frac{1}{30}=0.033$. 
 He  then point the laser pointer from a distance of about 1 meter at the
 apparent double slits in the coated side of the mirror and look at the 
 resulting light
 pattern on the white wall situated 1.75  meters from the mirror 
 (see Figure 5).  The obtained light  
 pattern  is expected to be  either that of  the interference type as in 
 Figure 1 in case the activated screen was one of the real double slit mirrors 
 or that of the 
 diffraction
 form of Figure 2  if this screen was the spurious one. 
  After obtaining the optical pattern upon the white wall
      the observer checks the operated screen  to see if it is one of the 
      real double-slit mirrors  or the faked one. 
        Thus,  if it is found to be  one of the true double-slit mirrors it is 
 returned to the pile of 30 mirrors from which another one (which may be the
 former) was randomely chosen for a new experiment of the type just described. 
 This repetition was stopped only  when the involved screen was found to be the
 spurious one. \par 
  Now, since  
  the chance of randomely picking the faked  screen is only $0.033$ 
 whereas that for chosing the real one is $0.97$  one, naturally,  
 have to
 repeat the described experiment a large number of times  until the spurious  
 screen
 was found. Thus,   this experiment was, actually,  repeated 228 times 
 over several
 weeks and for each of these experiments we have first obtained interference
  pattern and then found that the activated screen was one of the 29 real
  double-slit screens as expected.  
  The 229-th experiment  began, as its predecessors,
 without knowing and, therefore, without using the true nature of  the 
 involved screen and, as before, the
 obtained light pattern was of the interference type but upon looking closely 
 at
 the relevant screen it has, somewhat unexpectedly, turned out to be the 
 spurious one
 which is, actually, a single slit.   
 That is, 
 {\it interference pattern, which typically result from $n$-slit
  screen ($n \geq 2$) was,  actually,  obtained from single-slit one}. \par 
  In order to be sure of
  this  result a new series of these experiments was again
  repeated but this time each optical pattern is photographed by a digital
  camera \cite{camera} {\it before}
  checking  the true nature of the activated screen. 
   As before, hundreds of them (216) were performed over several weeks 
   before the
  faked double-slit screen were turned up at the 217-th experiment. 
  As   for the former series of experiments the finding of this actually
  single-slit screen was preceded by obtaining the $n$-slit interference 
  pattern ($n \geq 2$) 
  and not  the expected single-slit diffraction one.  This time, unlike the
  former series, all the obtained 217 optical patterns were photographed before
  checking the nature of the employed screen. A photograph of the optical
  pattern resulting  from the 217-th experiment   
  is shown in Figure 8 and  one may see that it is
 of the interference pattern kind as realized when comparing it to Figure 6 
 which
 shows the optical  pattern obtained from a real double-slit screen.  \par

\begin{figure}
\includegraphics[width=15cm]{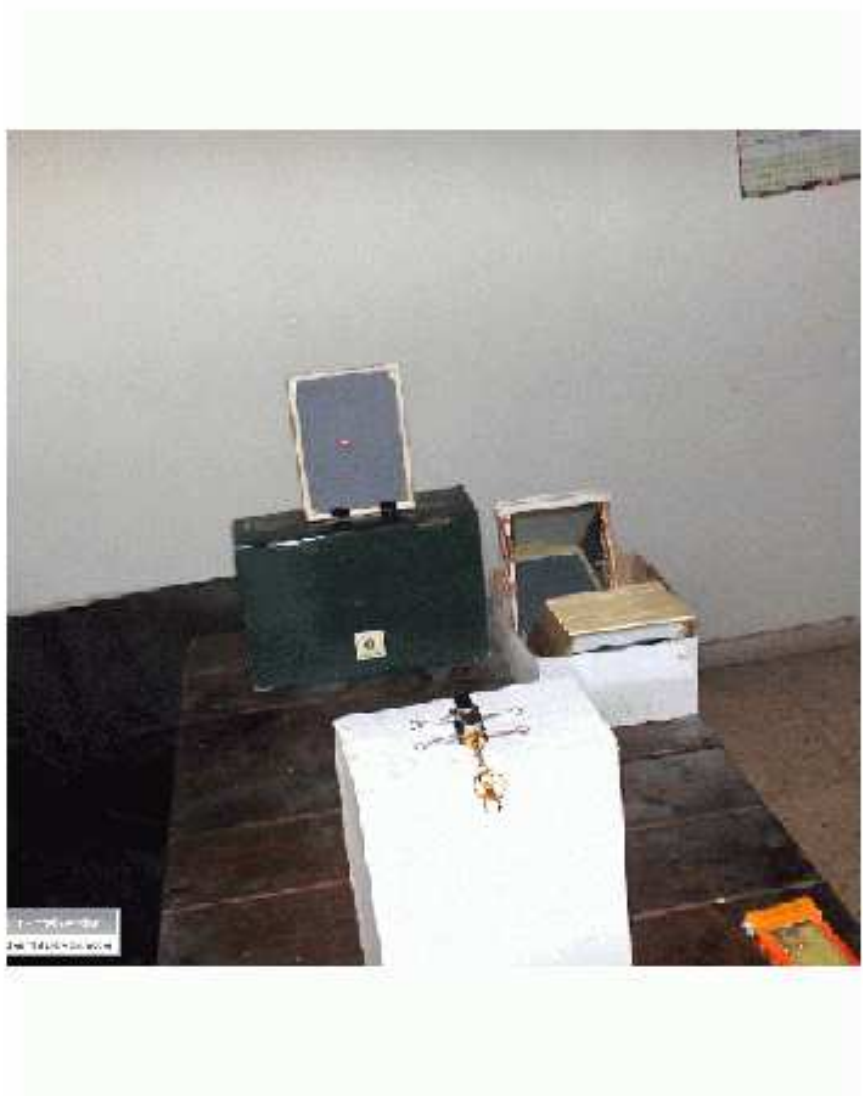}

     \caption{ This figure is a  digital camera photograph of the  arrangement used to 
     perform the
     experiment described in Section II. At the front 
     one may see on a white box the red laser
     pointer hold by two binding clips. At the back  one may see 
     on another box  the slitted mirror which is also hold by 
     another pair of 
     binding clips. The red laser ray from the pointer is shown at 
     the back of
     the mirror and with a larger resolution one may even discern the 
     optical
     pattern on the white wall. At the right one may see in 
     another box the
     ensemble of mirrors used.  }
     \end{figure}

     \begin{figure}
     \centerline{
\includegraphics[width=14cm]{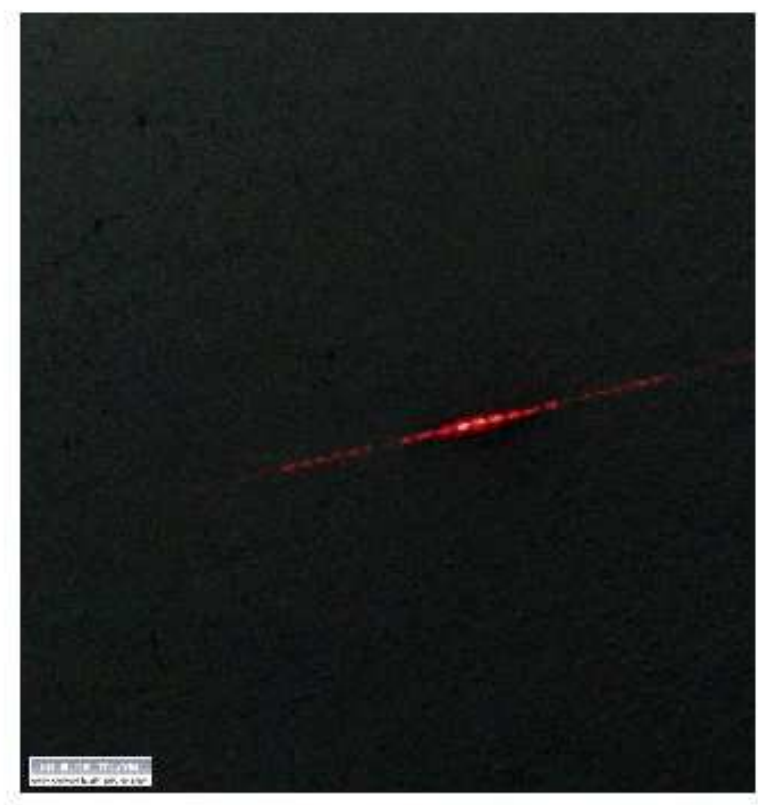}  }
\caption{The optical interference pattern obtained from one of the 
29 real
double-slit mirrors where the nature of this screen is known but not 
the data
about the routes through the slits. }
     \end{figure} 
  
\begin{figure}
\begin{minipage}{0.49\linewidth}
\includegraphics[width=11.5cm]{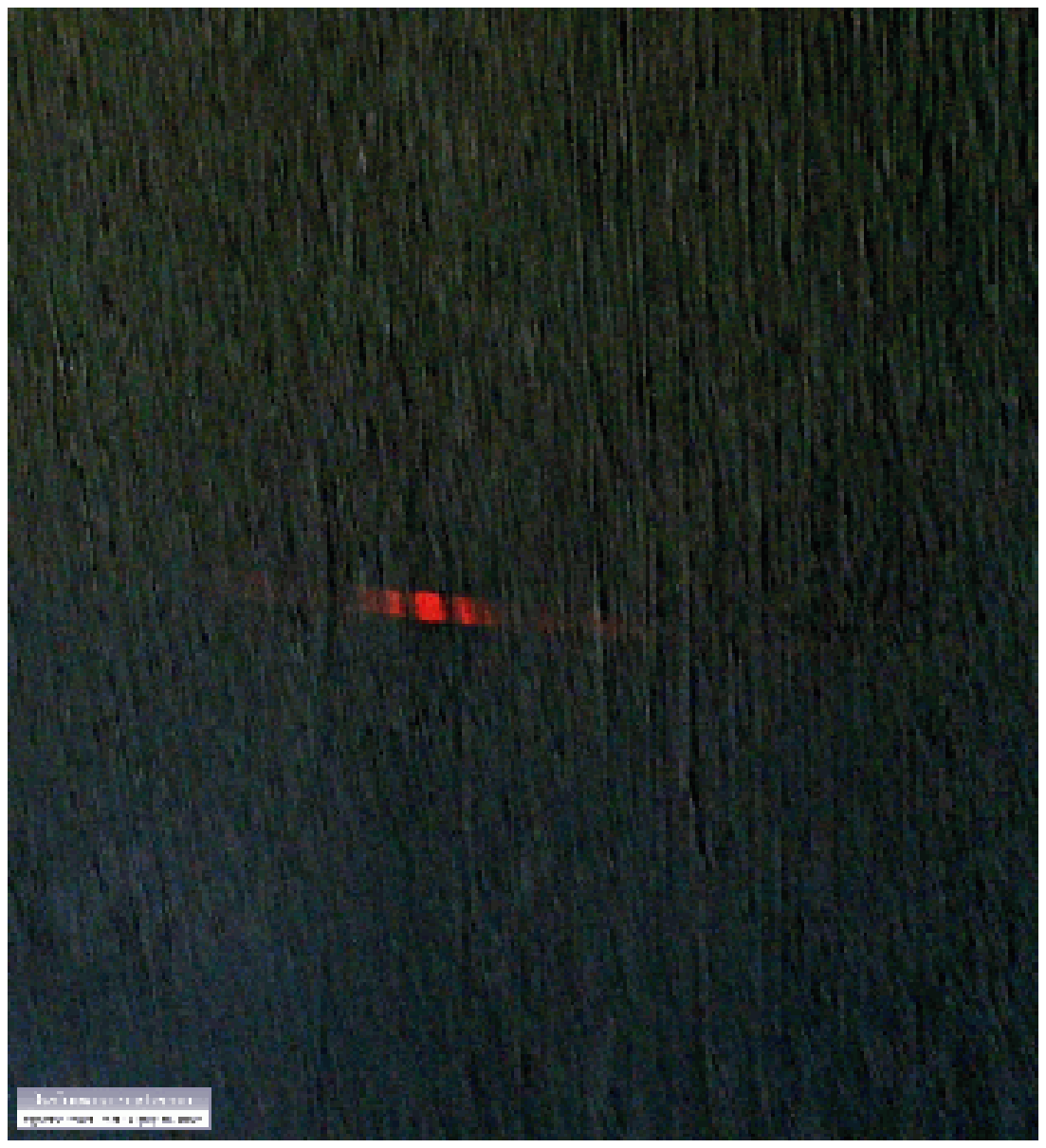}
\end{minipage} \hfill
\begin{minipage}{0.41\linewidth}
\includegraphics[width=11.5cm]{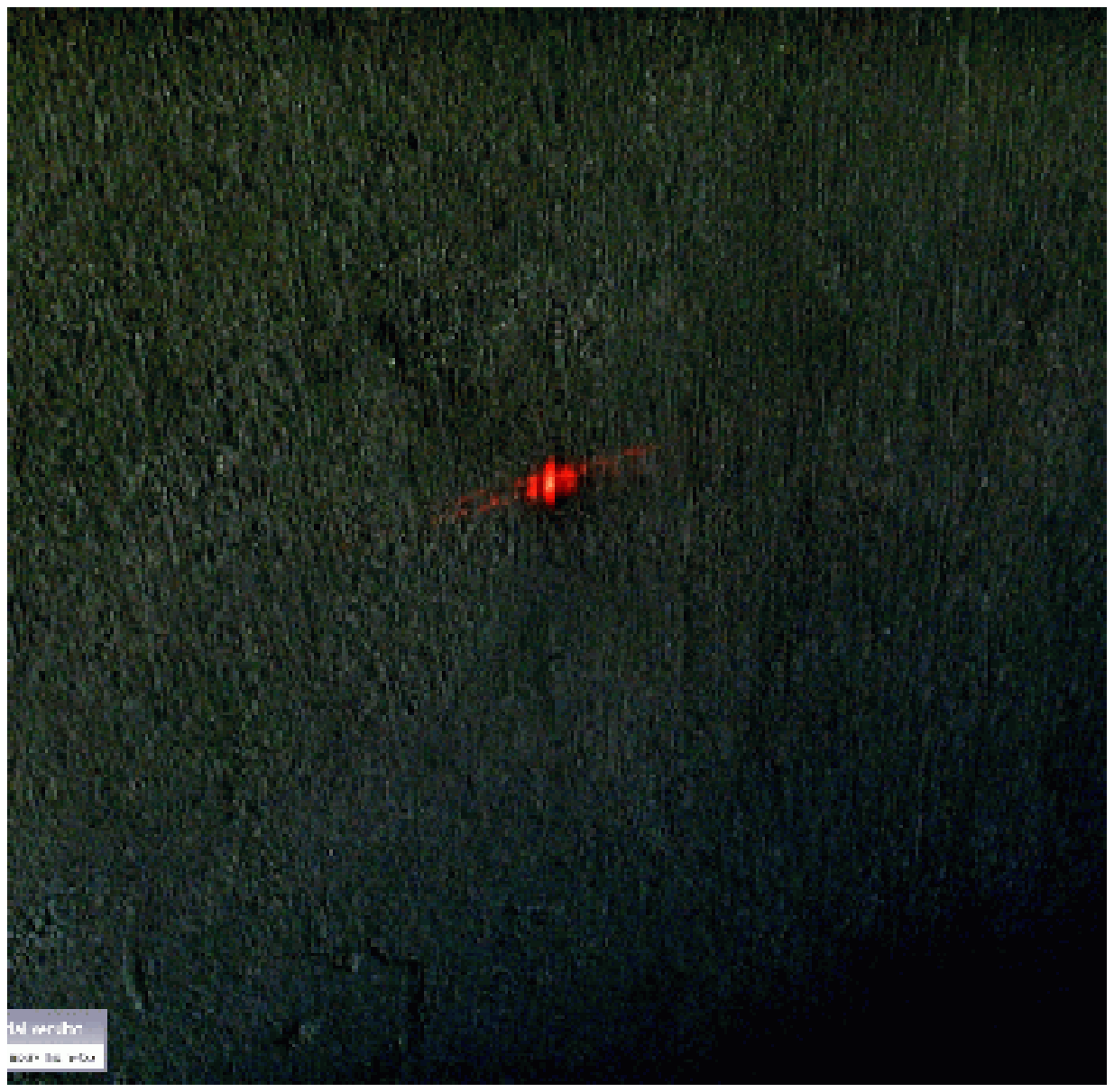}
\end{minipage}
 \caption{At  the left  and right Subfigures  we see  photographs 
  of the optical 
    single-slit diffraction 
    pattern taken at the start of the experiment  from the spurious 
    double-slit screen where the true nature of this screen 
    was known and used
    during the experiment.  The distance between the mirror and wall, 
     used for
    obtaining the optical pattern of the left picture,   was larger  
    compared to
    that used for the right picture. 
    }
     \end{figure}

 \begin{figure}
\centerline{
\includegraphics[width=14cm]{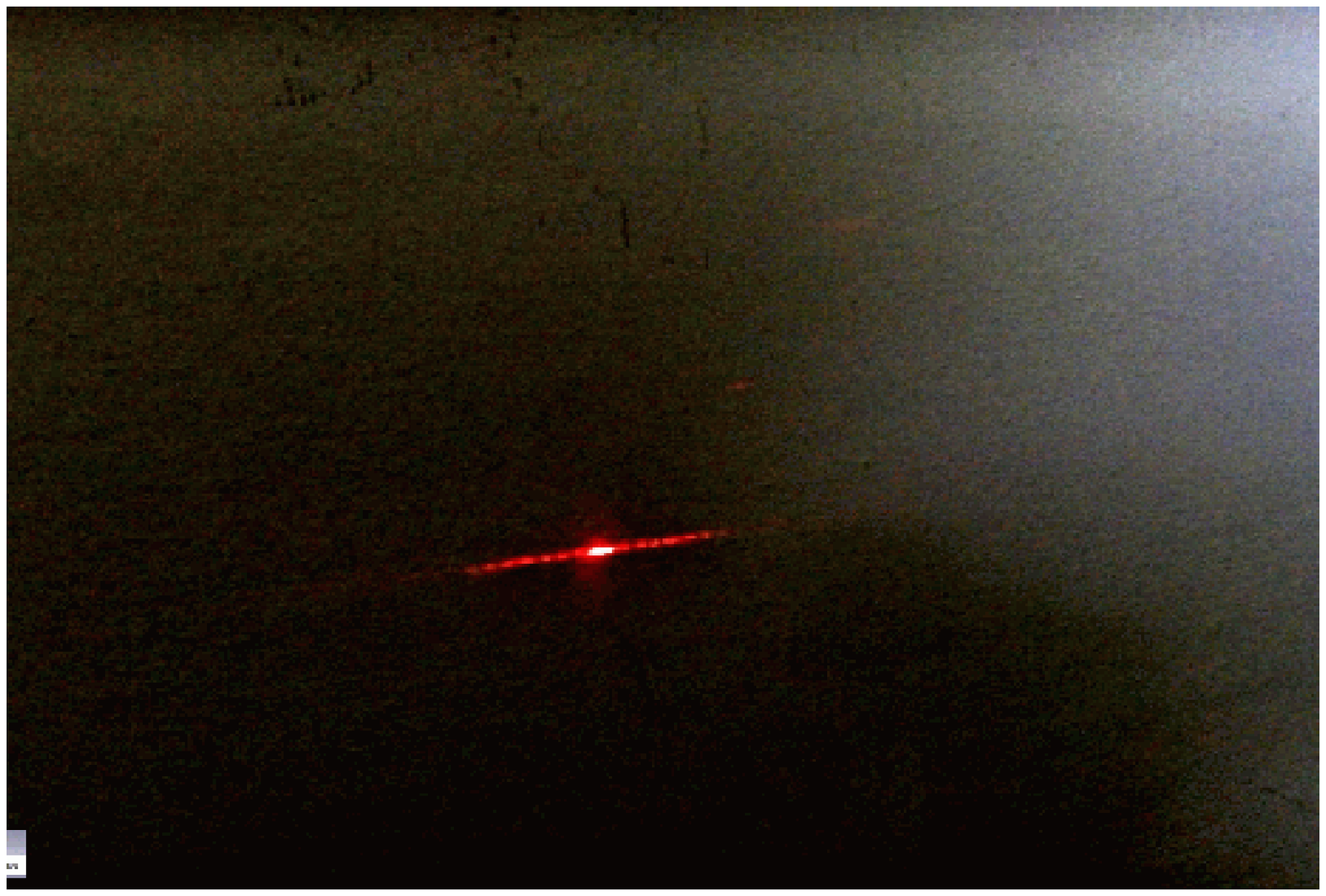}}
     
    \caption{This figure is a photograph of the optical  interference 
    pattern obtained  from the spurious 
    double-slit screen where the true nature of this screen 
    was not known and,
    therefore, not used
    during the experiment. Note the similarity of this pattern, 
    actually
    obtained from single-slit screen,  to that of Figure
    6 which was obtained from one of the real double-slit screens. 
    Comparing this 
    photograph to
    those of Figure 7, which are  obtained from the same screen,  
    one may realize
    that the use  (or not) during the experiment of the relevant data 
     is the important factor which
    determines the obtained optical pattern.  }
     \end{figure}

  Analysing the former results one may conclude  
    that performing these experiments under the  conditions of
    not knowing the nature of screen and, therefore, not using during the 
    experiments the data about 
    the  routes through the slits   changes  the  form  of these routes 
     from   the  diffraction pattern of  Figure 2, $b$   
     to the $n$-slit interference one ($n \geq 2$) shown in Figure 1 ,$b$.  
     This outcome 
     for the SSE 
     together with the   mentioned  
       results
      obtained   for any other $n$-slit  screen ($n \geq 2$)  led one to
      conclude that 
       interference pattern is obtained for any $n$-slit screen, 
       even for $n=1$,
         so long as the mentioned data were not 
     used during the experiment. When, however, these data are used during the
     experiment one obtains for such $n$-slit screen ($n \geq 1$) 
     $n$ diffraction patterns 
     each of them of the SSE kind shown in Figure 2. 
      In other words, considering  the central region
     around the order $m=0$,  one may realize that,
     when the noted data are not used during the experiments, this region
     becomes  
    striped and fringed  compared to the nonfringed form it has when these
    data are used.   That is, an interval between photon routes, 
     which were zero in the region around $m=0$ when the noted data were used
     (see Figures 2-4),  
      has been formed when these data were  not used (see 
     Figure 1).  We calculate in the following the length of this interval 
     which is shown in Figure 9 between  neighbouring maxima. 
      The path difference between  two
     rays from the two slits is shown at the left of Figure 9. Thus,  
      denoting by ${\cal B}$ and ${\cal D}$ the respective 
    bright and dark fringes upon
    the photosensitive 
     screen one may use Figure 9 and write           
     these path differences  
   for the DSE as  \cite{jenkins}

\begin{eqnarray}  && d\sin(w)_{{\cal B}}=
m\lambda \label{e1}  \\ 
 && d\sin(w)_{{\cal D}}=
(m+\frac{1}{2})\lambda  \nonumber 
\end{eqnarray}
 As shown in Figure 9,   $d$ and $d\sin(w)$  
 respectively denote the interval between the two slits and 
 the path difference between the two interfering waves.   By   
 $\lambda$ and $m=0, 1, 2,....$ we denote the wavelength of the light from
the source and   the order of interference.  For the single
slit experiment (SSE) one may use the following  expressions \cite{jenkins} 
\begin{eqnarray}  && b\sin(w)_{{\cal B}}=
m\lambda  \label{e2} \\ 
&& b\sin(w)_{{\cal D}}=
(m+\frac{1}{2})\lambda,  \nonumber 
\end{eqnarray}
where $b$ is the length of the slit and $b\sin(w)$ is the path difference
between rays diffracted from the ends of the slit. That is,   
 the locations of the  different maxima and minima do not result 
from any interference   but from 
diffraction through the single slit. As a result,  the maximum
intensity  for the order $m=0$  is greater 
by several order of magnitudes
from the corresponding maxima shown for the orders $m=1, 2, ...$ 
(see Figure 2) and from those of the DSE.  That is, most photons which 
diffract through the single slit
propagate parallel to each other in the forward straight direction which
explains the large intensity for the order $m=0$.   \par 
In Figure 9 we show  the two ordered maxima $m=0$ and $m=1$, denoted in 
the following
by $m_0$ and $m_1$,  and  calculate the interval between them  by
considering  the right angled triangle build from the sides
$m_0-x_1$, $g$ and $f$ where $x_1$ denotes the first minimum.  
From this triangle one  obtain $\tan(w)=\frac{(m_0-x_1)}{g}$ so that 
 the sought-for length $\Delta(w)=m_0-m_1$ is  
   \begin{equation} \label{e3}
  \Delta(w) = m_0-m_1= 2g\tan(w)
\end{equation} 
As noted, the  changed photon routes  
 may  be paralleled  to  
corresponding geodesics
changes. These changes are extensively discussed  
\cite{bergmann,mtw,hartle,hawking} 
 in the framework of general
relativity (see, especially, the annotated references in \cite{mtw}) 
 as resulting from corresponding spacetime changes.   The later
changes are, especially, tracked to GW's \cite{thorne} whose 
intrinsic spacetime 
geometry \cite{yurtsever} is imprinted upon the traversed spacetime. 
    As mentioned,   we pay special attention to the more appropriate 
    source-free GW's and   consider  the Brill's and plane GW's  
    (we may also discuss  
the Kuchar's cylindrical source-free 
GW's \cite{mtw,kuchar1}).   \par   
Thus, as for   the optical slitted-screens   in which one 
may either use during the
experiments the data about the
 routes through the slits (which result in  obtaining   
diffraction patterns)
 or not using these data (which result, 
 for any $n$-slit ecreen ($n \geq 1$), 
in obtaining   fringed interference pattern)   one may, likewise, 
discern two   
similar
gravitational states. One state may be characterized by some assumed spacetime 
metrics
such as the Brill's or the almost flat metrics which is used in discussing 
plane GW's. The second gravitational state is characterized by a metrics 
which, actually, is the fringing of   the former in a manner similar to 
the stripes of the interference
pattern which may be regarded as the fringing,  especially, in the 
neighbourhood of the
orders $m=0$,  of the diffraction patterns of Figures 2-4.  
 The  Brill's metrics and its corresponding nonfringed trapped 
surface \cite{eppley} is
reviewed in Appendix A and the  
almost flat metrics with the relevant   
nonfringed trapped surface \cite{bar1}
is represented in Appendix B.  The fringed 
  trapped surfaces 
 is discussed in the following section for the Brill's
GW and in Section IV for the almost flat plane GW's.

    \begin{figure}
\centerline{
\begin{turn}{-90}
\includegraphics[width=12cm]{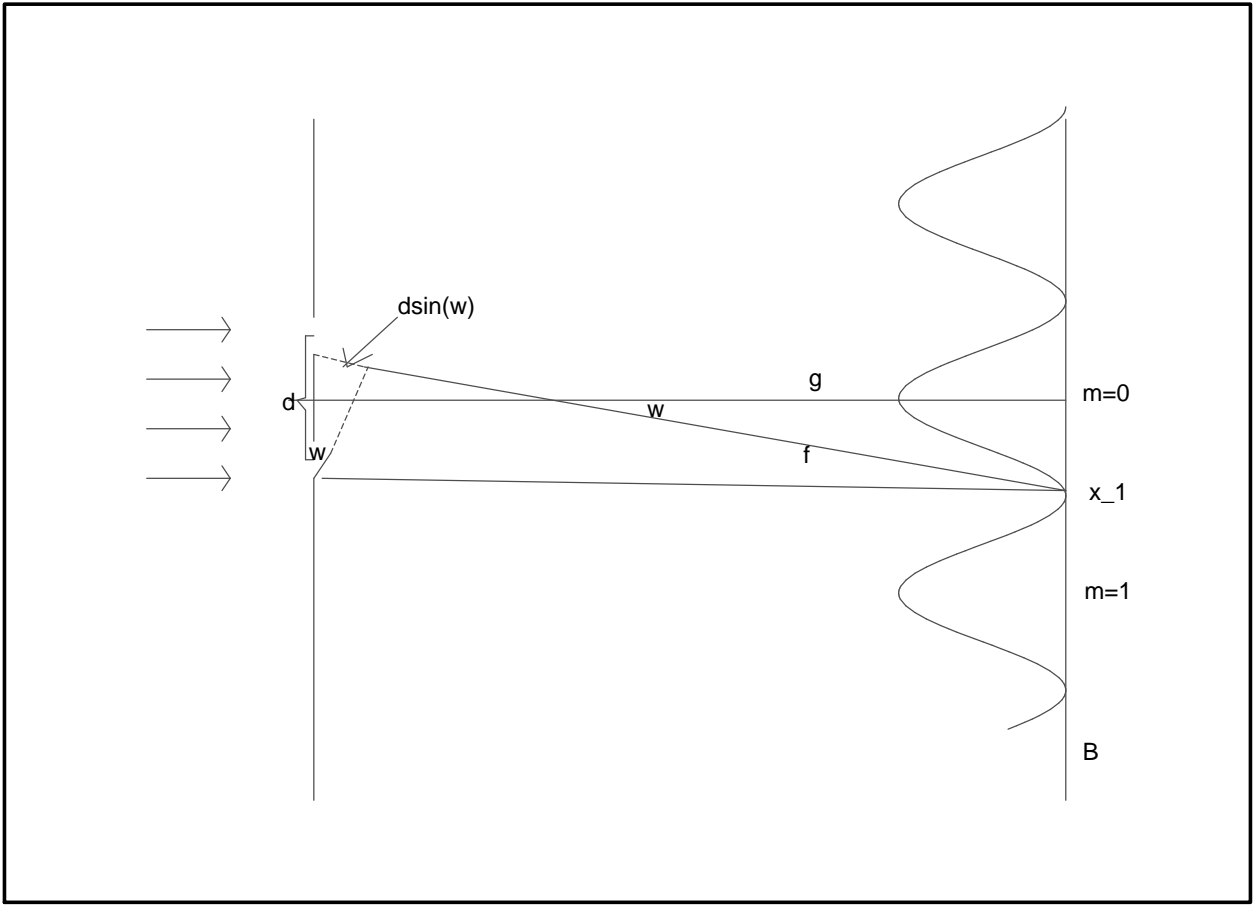}
\end{turn}}

\caption{A schematic representation of the double-slit array,  used 
to obtain
interference patterns, is shown. The path
difference $d\sin(\phi)$ and the sinusoidal interference form 
are shown. Using
this figure one may obtain Eq (\ref{e3}) for the interval between 
the orders
$m=0$ and $m=1$. }
\end{figure}

\markright{THE FRINGED TRAPPED SURFACE RESULTING FROM THE BRILL GW's}

 \protect  \section{The fringed trapped surface resulting from the Brill GW's} 
The source-free Brill GW's are discussed  in the framework
of the ADM formalism \cite{mtw,adm}. In this canonical
formulation of general relativity one, generally,  consider the
simplified case of time and axial symmetries and no rotation
\cite{brill1,eppley,gentle}. Under these
conditions  one  finds a solution \cite{brill1,eppley,gentle} to the 
Einstein vacuum field equations
which represents, as mentioned, pure source-free gravitational wave with
positive energy \cite{brill1,eppley,gentle,alcubierre}. A short review of 
the ADM
canonical theory \cite{mtw} with the former conditions is represented in 
Appendix $A$.
 Strong Brill  GW's are involved with the appearance of a
marginally trapped surfaces which are equivalent, for the time-symmetric condition
discussed here, to  minimal area surfaces  \cite{eppley}. These surfaces
may be represented in a cartesian coordinates by using embedding diagrams of
them
\cite{mtw,eppley}. As known \cite{mtw,eppley}, to embed the whole surface is
difficult so one, generally, resort to the task of embedding a plane through the
equator which is simpler due to the assumed rotational symmetry. \par
As mentioned, we have obtained for any $n$-slit screen,  even for 
$n = 1$ as   
 described  in Section II,  
an interference pattern which are 
periodically alternating sequence of light and dark bands (see 
 Figure 1, $b$) where the bright bands allow photons to pass along them and the
 dark ones do not allow them.  Using the equivalence principle these 
  alternating fringes may, as mentioned, be
 discussed  as periodically alternating bands of geodesics. That is,  
  the optical 
 bright bands correspond to strong curvature \cite{bergmann,mtw,hartle} 
 geodesics and the optical 
 dark bands
  to  the weak
  curvature ones.   As noted, these alternating bands of strong and weak
  curvatures may be considered as 
 forming   a  trapped fringed surface which, theoretically,  can be 
 embedded in
 an Euclidean space \cite{mtw,eppley,brill2}.   The embedding
procedure may, analytically, be expressed by requiring the metric of the
 equator \cite{eppley} to be equal to that of a rotation  surface (in Euclidean 
 space) 
which is formed from a periodic alternating allowed and disallowed 
bands. That
is, one may write 
\begin{eqnarray} && x^B=F^B(\rho,\phi)\cdot \cos(\phi) \nonumber \\
&&  y^B=F^B(\rho,\phi)\cdot \sin(\phi) \label{e4} \\
&& z^B=h^B(\rho),   \nonumber \end{eqnarray}  
where the function $h^B(\rho)$ does not depend upon the variable $\phi$ and 
the superscript $B$ denote that we refer to the Brill source-free case. 
The function $F^B(\rho,\phi)$  may  be written as the following products
\begin{equation}  F^B(\rho,\phi)=f^B(\rho)\cdot M^B(\phi),  
  \label{e5}  \end{equation}
where the function $M^B(\phi)$ is introduced to ensure the mentioned periodic 
fringing 
and is defined as 
\begin{equation}  \label{e6} M^B(\phi) = \left\{ \begin{array}{ll}  1 & \ \ {\rm for} 
-\pi \le \phi \le -\frac{\pi}{2}  \\
0 &  \ \ {\rm for} 
-\frac{\pi}{2} < \phi < \frac{\pi}{2} \\ 1  & \ \ {\rm for}  
\frac{\pi}{2} \le \phi \le \pi  \end{array} \right. \end{equation} 
As seen,  the periodic function $M^B(\phi)$, which is shown in Figure 10,  
is piecewise monotonic and bounded on the
interval $(-\pi, \pi)$ and so it can be expanded in a Fourier series 
\cite{spiegel1,pipes}. Thus, using the Fourier analysis 
\cite{spiegel1,pipes} one may determine
the  Fourier coefficients as 
\begin{eqnarray} &&
a^B_0=\frac{1}{\pi}\int_{-\pi}^{\pi}M^B(\phi)d\phi=
\frac{1}{\pi}\bigl( \int_{-\pi}^{-\frac{\pi}{2}} 1 d\phi+ \int_{-\frac{\pi}{2}}
^{\frac{\pi}{2}} 0 d\phi+\int_{\frac{\pi}{2}}^{\pi} 1 d\phi
\bigr)=\frac{1}{\pi}\pi=1 \nonumber \\ && a^B_k=
\frac{1}{\pi}\int_{-\pi}^{\pi}M^B(\phi)\cos(k\phi)d\phi=
 \frac{1}{\pi}\int_{-\pi}^{-\frac{\pi}{2}} 1 \cos(k\phi)
 d\phi+  \nonumber  \\  &&  + \frac{1}{\pi}\int_{\frac{\pi}{2}}^{\pi} 1 \cos(k\phi)
 d\phi  = \frac{1}{\pi}\bigl(\frac{\sin(k\phi)}{k}\bigr)
  {\Large |}^{-\frac{\pi}{2}}
 _{-\pi}+\frac{1}{\pi}\bigl(\frac{\sin(k\phi)}{k}\bigr)
  {\Large |}^{\pi}
 _{\frac{\pi}{2}}= \nonumber  \\  && = 
- \frac{2}{\pi k}\sin(k\frac{\pi}{2})=  
 \left\{ \begin{array}{ll} 0 &  {\rm for \ k \ even} \\ -\frac{2}{\pi k} & 
 {\rm for \ k \ odd \ and \ k=4n-3 } \\ \frac{2}{\pi k} & 
 {\rm for \ k \ odd \ and \ k=4n-1 } \end{array} \right.   \label{e7}  
 \\  
  &&b^B_k=\frac{1}{\pi}\int_{-\pi}^{\pi}M^B(\phi)\sin(k\phi)d\phi=
\frac{1}{\pi} \int_{-\pi}^{-\frac{\pi}{2}} 1 \sin(k\phi)
 d\phi+\frac{1}{\pi} \int_{\frac{\pi}{2}}^{\pi} 1 \sin(k\phi)
 d\phi = \nonumber \\ && = -\frac{1}{\pi}\bigl(\frac{\cos(k\phi)}{k}\bigr) 
 {\Large |}^{-\frac{\pi}{2}}
 _{-\pi} -\frac{1}{\pi}\bigl(\frac{\cos(k\phi)}{k}\bigr) 
 {\Large |}^{\pi}
 _{\frac{\pi}{2}} =0,  
    \nonumber \end{eqnarray}
    where $n=1, \ 2, \ 3, \ 4......$. 
    From the last Eqs (\ref{e6})-(\ref{e7}) one may write the function
    $M^B(\phi)$ as 
    \begin{eqnarray} &&  M^B(\phi)=1-\frac{2}{\pi}
   \left(
   \frac{\cos(1\phi)}{1}+\frac{\cos(5\phi)}{5}+\ldots +
  \frac{\cos((4n-3)\phi)}{(4n-3)}+ \ldots \right)+ \nonumber \\ && +
  \frac{2}{\pi} \left( \frac{\cos(3\phi)}{3} +\frac{\cos(7\phi)}{7}
+\dots + \frac{\cos((4n-1)\phi)}{(4n-1)} +\ldots \right) =
\label{e8} \\ && = 1-\frac{2}{\pi} \left(
\sum_{n=1}^{\infty}\frac{\cos((4n-3)\phi)}{(4n-3)}- 
\sum_{n=1}^{\infty}\frac{\cos((4n-1)\phi)}{(4n-1)}
\right) \nonumber  
  \end{eqnarray}

\begin{figure}
\centerline{
\begin{turn}{-90}
\includegraphics[width=9cm]{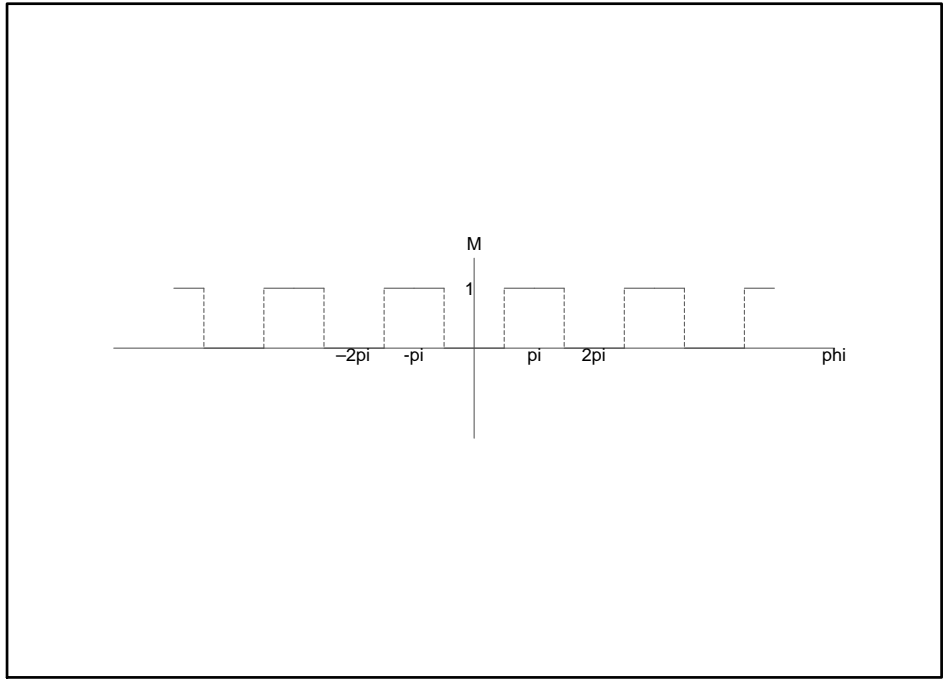}
\end{turn}}

     \caption{In this figure one may see the periodic 
     function $M(\phi)$ from Eq (\ref{e6}).  }
     \end{figure}
  
 Now,   using Eqs (\ref{e4})-(\ref{e5})  
and following the
discussion in \cite{eppley} one may  write for the metric on the equator 
\begin{eqnarray} &&
(ds^B)^2=(dx^B)^2+(dy^B)^2+(dz^B)^2=
\biggl((M^B)^2(\phi)(f^B)_{\rho}^2(\rho)+
\label{e9}  \\ && + (h^B)_{\rho}^2(\rho)\biggr)
d\rho^2+ (f^B)^2(\rho)\biggl(M^B)_{\phi}^2(\phi)+ 
(M^B)^2(\phi)\biggr)d\phi^2+ \nonumber \\ &&  
 +4M^B(\phi)M^B_{\phi}(\phi)f^B_{\rho}(\rho)f^B(\rho)
 d\rho d\phi,  \nonumber \end{eqnarray}
 where    $f^B_{\rho}(\rho),\ h^B_{\rho}(\rho)$  
 denote the respective 
derivatives of $f^B(\rho), \ h^B(\rho)$ with respect to $\rho$ and 
$M^B_{\phi}(\phi)$
is the derivative of $M^B(\phi)$  with respect to $\phi$.
 The  metrics from Eq (\ref{e9}) is equated to the Brill's one
 \cite{brill1,gentle,eppley} which is represented  in Appendix A.  
Note that  for performing the mentioned embedding one consider \cite{eppley} 
   only  the metrics on the equator \cite{eppley} 
 which is equated to the
 surface of rotation from  Eqs (\ref{e4}) and  also  the metrics for 
 the Brill
 GW's  is generally assumed for the norotation case  \cite{eppley} (see Appendix
 A).  Thus, one obtains from this  equating process
\begin{eqnarray} &&
(ds^B)^2=(dx^B)^2+(dy^B)^2+(dz^B)^2=
\biggl((M^B)^2(\phi)(f^B)_{\rho}^2(\rho)+
\nonumber  
\\  && + (h^B)_{\rho}^2(\rho)\biggr)
d\rho^2+   
f^2(\rho)\biggl((M^B)_{\phi}^2(\phi) +  
(M^B)^2(\phi)\biggr)d\phi^2= \label{e10} \\ 
 && = g^B_{\rho\rho}d\rho^2+g^B_{\phi\phi}d\phi^2= 
\psi^4e^{2Aq}d\rho^2+\psi^4\rho^2d\phi^2, \nonumber 
\end{eqnarray} 

where $g^B_{\rho\rho}$ and $g^B_{\phi\phi}$  are the $(\rho\rho)$ and
$(\phi\phi)$ components of the Brill metric tensor.   
   For  obtaining 
 $F^B(\rho,\phi)$, $F^B_{\rho}(\rho,\phi)$
and $h^B(\rho)$ one equates the coefficients of $d\rho^2$ and $d\phi^2$ as
follows 
\begin{eqnarray} &&
F^B(\rho,\phi) =f^B(\rho)M^B(\phi)=
\frac{\rho\psi^2}{\sqrt{1+\frac{(M^B)_{\phi}^2(\phi)}{(M^B)^2(\phi)}}} 
\nonumber \\ && 
F^B_{\rho}(\rho,\phi)=f^B(\rho)_{\rho}M^B(\phi)=
  \frac{1}{\sqrt{1+\frac{(M^B)_{\phi}^2(\phi)}{(M^B)^2(\phi)}}}\left(\psi^2+
  2\rho\psi\psi_{\rho}\right) 
   \label{e11} \\
&& h^B(\rho)=\int
d\rho\biggl(\psi^4e^{2Aq}-(F^B)_{\rho}^2(\rho,\phi)\biggr)^{\frac{1}{2}}=\int
d\rho\biggl(\psi^4e^{2Aq}-(M^B)^2(\phi)(f^B)_{\rho}^2(\rho)\biggr)^{\frac{1}{2}}
\nonumber 
\end{eqnarray}
The last expressions for $F^B(\rho,\phi)$, $F^B_{\rho}(\rho,\phi)$ and  
$h^B(\rho)$ define the embedded fringed surface $z(x,y)$ on the equator 
which has
the same geometry as that of the time-symmetric Brill's GW. These 
surfaces have the same general forms as those shown in Figure 11 for the
nonfringed surfaces of Eqs (\ref{$A_{13}$}) except that now, for appropriately
representing Eqs (\ref{e11}), these surfaces should be periodically fringed.
\par 
Now, as done in Section II  
 regarding the orders $m$   
of the fringed interference pattern
  (see Eqs (\ref{e1})-(\ref{e3}))  and the intervals between 
  neighbouring maxima   we also find  here the orders and 
  intervals related to   the fringed  trapped surface geometry   
   resulting from the
 Brill  GW's.  That is, realizing from Eq (\ref{e6}) and Figure 10 that 
 any two neighbouring  maximal  or minimal bands,  for which one respectively   
 have    $M^B(\phi)=1$  and 
 $M^B(\phi)=0$,      
  are separated by intervals of $\pm\pi$ 
 one may
 denote the angles related to these ordered  
  bands by $\phi_{m}=\phi_0 \pm m\pi$ where $\phi_0$ corresponds to $m=0$.  
  Note that for $m$ even (positive or
  negative) one always have 
 $ \cos(\phi_{m})=\cos(\phi_0)$, $ \sin(\phi_{m})=\sin(\phi_0)$ and for $m$ odd
 (positive or negative) 
 $ \cos(\phi_{m})=-\cos(\phi_0)$, $ \sin(\phi_{m})=-\sin(\phi_0)$. Also, one
always  have for any $m$,  positive or negative,  even or odd,  
 $ \cos(\phi_{m+1})=-\cos(\phi_m)$, $ \sin(\phi_{m+1})=-\sin(\phi_m)$. 
   Thus,    
   remembering that the  coordinates of the fringed trapped 
   surfaces are given by Eqs (\ref{e4})  
    one may write    in correspondence with Eqs  (\ref{e1})-(\ref{e2}) 
    \begin{eqnarray}  && x^B_{m}=
F^B(\rho,\phi_m)\cos(\phi_m) =f^B(\rho)M^B(\phi_m)\cos(\phi_m) \label{e12} \\ 
 && y^B_{m}=
F^B(\rho,\phi_m)\sin(\phi_m)=f^B(\rho)M^B(\phi_m)\sin(\phi_m), \nonumber  
\end{eqnarray} 
where for maximal bands  one should have either  
 $-\pi  \le \phi \le -\frac{\pi}{2} $
 or 
$\frac{\pi}{2}  \le \phi \le \pi $ for which $M^B(\phi)=1$ (see Eq (\ref{e6})) and 
for minimal bands   $\phi$ should be from the range 
$-\frac{\pi}{2}  < \phi < \frac{\pi}{2}$ for which  $M^B(\phi)=0$ (see Eq
(\ref{e6})).   
 Now, denoting the interval between two neighbouring $m$ orders 
  in the fringed trapped surface
geometry  by $\Delta^B(\rho)$ and using the relations 
$\cos(\phi_{(m+1)})=-\cos(\phi_m)$,  $\sin(\phi_{(m+1)})=-\sin(\phi_m)$ and  
$M^B(\phi_{(m+1)})+M^B(\phi_m)=2$   
(see  Eq (\ref{e8})) 
one may write in correspondence with Eq (\ref{e3})    
\begin{eqnarray}   && 
\Delta^B(\rho)= \sqrt{\left(x^B_{(m+1)}-x^B_{m}\right)^2+
\left(y^B_{(m+1)}-y^B_{m}\right)^2}= \nonumber \\ && = 
\biggl\{ \biggl(f^B(\rho)M^B(\phi_{m+1)})\cos(\phi_{(m+1)})-
f^B(\rho)M^B(\phi_m)\cos(\phi_m)\biggr)^2+ \label{e13} \\ && +
\biggl(f^B(\rho)M^B(\phi_{m+1)})\sin(\phi_{(m+1)})-
f^B(\rho)M^B(\phi_m)\sin(\phi_m)\biggr)^2\biggr\}^{\frac{1}{2}}= \nonumber \\ 
&& = 
\biggl\{(f^B)^2(\rho)\biggl(M^B(\phi_{(m+1)})+
M^B(\phi_m)\biggr)^2\biggl(\cos^2(\phi_m)+\sin^2(\phi_m)\biggr)\biggr\}^{\frac{1}{2}}
=    \nonumber \\ && = 2f^B(\rho) = 
\frac{2\rho\psi^2}{\sqrt{((M^B)^2(\phi)+(M^B)_{\phi}^2(\phi))}},   \nonumber
\end{eqnarray}
 where the last result follows from  the first of Eqs (\ref{e11}).

 \markright{THE FRINGED TRAPPED SURFACE RESULTING FROM THE LINEARIZED.....}
 
  \section{The fringed trapped surface resulting from the linearized plane GW's}

    We, now,  find, as for the Brill   GW's discussed in the former section, 
       the required fringed trapped   surface related to the linearized plane
       GW's and begin by 
assuming, as in \cite{eppley}, that its metric is 
 that of a surface of rotation $z(x,y)$ related to Euclidean space. 
 That is, one may write 
 \begin{eqnarray} && x^P=F^P(\rho,\phi)\cos(\phi)=
 f^P(\rho)M^P(\phi)\cos(\phi), \nonumber  \\ 
&& y^P=F^P(\rho,\phi)\sin(\phi)=f^P(\rho)M^P(\phi)\sin(\phi), 
\label{e14}  \\ 
&&  z^P=h^P(\rho), 
 \nonumber \end{eqnarray} 
 where we have superscripted   the quantities $F(\rho,\phi)$, $M(\phi)$, 
$f(\rho)$ and $h(\rho)$  by 
$P$ to emphasize that they denote now plane GW's. 
We have also, as for the Brill case in Section III and for the same reason of
ensuring the periodic fringing,    
introduce the function
$M^P(\phi)$ which is identical to the $M^B(\phi)$ from Eq (\ref{e6}). 
Thus,   one may use the piecewise monotony 
and boundness of  $M^P(\phi)$ and expand it in a Fourier series
\cite{spiegel1,pipes} 
for  
obtaining   the appropriate coefficients $a^P_0, \ a^P_k, \ b^P_k$ which are 
identical 
to the $a^B_0, \ a^B_k, \ b^B_k$ from Eqs (\ref{e7}).   
  One may, also,  obtain the following expression for the metrics 
(compare with the Brill's case of Eq
 (\ref{e10}))  
\begin{eqnarray} &&
(ds^P)^2=(dx^P)^2+(dy^P)^2+(dz^P)^2=\biggl((M^P)^2(\phi)
(f^P)_{\rho}^2(\rho)+
\label{e15} \\ && +(h^P)_{\rho}^2(\rho)\biggr)
d\rho^2 + 
(f^P)^2(\rho)\biggl( (M^P)_{\phi}^2(\phi)+ 
(M^P)^2(\phi)\biggr)d\phi^2 + \nonumber  \\ && 
 +4M^P(\phi)M^P_{\phi}(\phi)f^P_{\rho}(\rho)f^P(\rho)
 d\rho
 d\phi,  \nonumber \end{eqnarray}
 where    $f^P_{\rho}(\rho),\ h^P_{\rho}(\rho)$  
 denote the respective 
derivatives of $f^P(\rho), \ h^P(\rho)$ with respect to $\rho$ and 
$M^P_{\phi}(\phi)$
is the derivative of $M^P(\phi)$  with respect to $\phi$.
The  metrics from Eq (\ref{e15}) is equated to that obtained in 
Eq (\ref{$B_{18}$}) in Appendix B where,   
  as for the Brill case discussed in Section III, 
  one  considers only  the metrics on the equator \cite{eppley} which is 
 equated to the
 surface of rotation from  Eqs (\ref{e14}) and also assume  the norotation case. 
  Thus, using the expressions for 
 $h^{TT}_{{\hat {\bf \rho}}{\hat{\bf \rho}}}$ 
 and $h^{TT}_{{\hat {\bf \phi}}{\hat{\bf \phi}}}$
 from Eq (\ref{$B_{18}$}) in Appendix B 
  one may write the metric of the fringed trapped surface as  
\begin{eqnarray} &&
(ds^P)^2=(dx^P)^2+(dy^P)^2=
\biggl((M^P)^2(\phi)(f^P)^2_{\rho}(\rho)+(h^P)_{\rho}^2(\rho)\biggr)
d\rho^2+  \nonumber  \\ && + 
(f^P)^2(\rho)\biggl((M^P)_{\phi}^2(\phi) +  
(M^P)^2(\phi)\biggr)d\phi^2=  
 h^{TT}_{{\hat {\bf \rho}}{\hat{\bf \rho}}}d^2\rho +
h^{TT}_{{\hat {\bf \phi}}{\hat{\bf \phi}}}d^2\phi  = \nonumber \\ && = 
\cos(kz -ft)\biggl[\frac{\sin(4\phi)}{2}(A_{\times}-A_+)
\left({\bf e}_{\hat {\bf \rho}}\otimes 
{\bf e}_{\hat {\bf \phi}}+{\bf e}_{\hat {\bf \phi}}\otimes 
{\bf e}_{\hat {\bf \rho}}\right)+\label{e16}  \\ && + 
\left(A_+\cos^2(2\phi)+A_{\times}\sin^2(2\phi)\right)
 \biggl( {\bf e}_{\hat {\bf \rho}}\otimes 
{\bf e}_{\hat {\bf
\rho}}-{\bf e}_{\hat {\bf \phi}}\otimes 
{\bf e}_{\hat {\bf
\phi}}\biggr)\biggr]
 \left(d^2\rho-\rho^2d\phi^2\right) \nonumber 
 \end{eqnarray}
 
 The appropriate  expressions  for 
$F^P(\rho,\phi)=f^P(\rho)M^P(\phi)$  and  $h^P(\rho)$ 
from Eq (\ref{e14}) which
 determine the intrinsic geometry of the fringed trapped surface 
 are obtained from Eq (\ref{e16})  by equating the respective coefficients  
 of both $d\rho^2$ and
 $d\phi^2$ as follows
 \begin{eqnarray} 
&&  (M^P)^2(\phi)(f^P)^2_{\rho}(\rho)+(h^P)_{\rho}^2(\rho)=
\cos(kz -ft) \cdot  \label{e17} \\ && \cdot  
\biggl[\frac{\sin(4\phi)}{2}(A_{\times}-A_+) 
 \cdot 
\left({\bf e}_{\hat {\bf \rho}}\otimes 
{\bf e}_{\hat {\bf \phi}}+{\bf e}_{\hat {\bf \phi}}\otimes 
{\bf e}_{\hat {\bf \rho}}\right) + 
\biggl(A_+\cos^2(2\phi)+  \nonumber \\ && + A_{\times}\sin^2(2\phi)\biggr)
 \biggl( {\bf e}_{\hat {\bf \rho}}\otimes 
{\bf e}_{\hat {\bf
\rho}}-{\bf e}_{\hat {\bf \phi}}\otimes 
{\bf e}_{\hat {\bf
\phi}}\biggr)\biggr]  \nonumber  
\end{eqnarray} 
Note that, as emphasized after Eqs (\ref{$B_{17}$}) in Appendix $B$, 
the expressions 
${\bf e}_{\hat {\bf \rho}}\otimes 
{\bf e}_{\hat {\bf \rho}}$,  ${\bf e}_{\hat {\bf \rho}}\otimes 
{\bf e}_{\hat {\bf \phi}}$ and ${\bf e}_{\hat {\bf \phi}}\otimes 
{\bf e}_{\hat {\bf \phi}}$ are tensor components in the $\rho\rho$, 
$\rho\phi$   and $\phi\phi$  directions and so, of course,  
they are not tensors proper.  This is of course the generalization  of the
components of some space vector, such as the $x$, $y$ and $z$ components of it, 
which although may have functional properties they certainly are not vectors.  
  Thus,   in the last equation and the 
 following ones we have compared these tensor components  
  to functions and even   took 
their  square roots.

\begin{eqnarray} 
&& (f^P)^2(\rho)\biggl((M^P)_{\phi}^2(\phi) +  
(M^P)^2(\phi)\biggr)=
\rho^2\cos(kz -ft)  \cdot \label{e18}  \\ && \cdot 
\biggl[\frac{\sin(4\phi)}{2}(A_+-A_{\times}) 
 \cdot 
\left({\bf e}_{\hat {\bf \rho}}\otimes 
{\bf e}_{\hat {\bf \phi}}+{\bf e}_{\hat {\bf \phi}}\otimes 
{\bf e}_{\hat {\bf \rho}}\right) + 
\biggl(A_+\cos^2(2\phi)+  \nonumber \\ && + A_{\times}\sin^2(2\phi)\biggr)
 \biggl( {\bf e}_{\hat {\bf \phi}}\otimes 
{\bf e}_{\hat {\bf
\phi}}-{\bf e}_{\hat {\bf \rho}}\otimes 
{\bf e}_{\hat {\bf
\rho}}\biggr)\biggr] \nonumber 
\end{eqnarray}
From Eq (\ref{e18}) one obtains for $F^P(\rho,\phi)$
\begin{eqnarray} 
&& F^P(\rho,\phi)=f^P(\rho)M^P(\phi)=
\biggl\{ \left(\frac{1}{1+\frac{(M^P)_{\phi}^2(\phi)}{(M^P)^2(\phi)}}\right)
\rho^2\cos(kz -ft) \cdot \label{e19} \\ && \cdot 
\biggl[\frac{\sin(4\phi)}{2}(A_+-A_{\times}) 
 \cdot 
\left({\bf e}_{\hat {\bf \rho}}\otimes 
{\bf e}_{\hat {\bf \phi}}+{\bf e}_{\hat {\bf \phi}}\otimes 
{\bf e}_{\hat {\bf \rho}}\right) + 
\biggl(A_+\cos^2(2\phi)+ \nonumber \\ 
 && + A_{\times}\sin^2(2\phi)\biggr)
 \biggl( {\bf e}_{\hat {\bf \phi}}\otimes 
{\bf e}_{\hat {\bf
\phi}}-{\bf e}_{\hat {\bf \rho}}\otimes 
{\bf e}_{\hat {\bf
\rho}}\biggr)\biggr] \biggr \}^{\frac{1}{2}} 
\nonumber \end{eqnarray}
From the last equation one obtains for the derivative of 
$F^P(\rho,\phi)$ with respect to $\rho$
\begin{eqnarray} 
&& F^P_{\rho}(\rho,\phi)=f^P_{\rho}(\rho)M^P(\phi)=
\biggl\{ \left(\frac{1}{1+\frac{(M^P)_{\phi}^2(\phi)}{(M^P)^2(\phi)}}\right)
\cos(kz -ft) \cdot \label{e20}  \\ && \cdot 
\biggl[\frac{\sin(4\phi)}{2}(A_+-A_{\times})  
\cdot 
\left({\bf e}_{\hat {\bf \rho}}\otimes 
{\bf e}_{\hat {\bf \phi}}+{\bf e}_{\hat {\bf \phi}}\otimes 
{\bf e}_{\hat {\bf \rho}}\right) + 
\biggl(A_+\cos^2(2\phi)+  \nonumber \\ && + A_{\times}\sin^2(2\phi)\biggr)
 \biggl( {\bf e}_{\hat {\bf \phi}}\otimes 
{\bf e}_{\hat {\bf
\phi}}-{\bf e}_{\hat {\bf \rho}}\otimes 
{\bf e}_{\hat {\bf
\rho}}\biggr)\biggr] \biggr \}^{\frac{1}{2}} 
\nonumber \end{eqnarray}
And, using Eq (\ref{e17}), one may obtain for 
  $h^P(\rho)$ 
  \begin{eqnarray} && 
 h^P(\rho)=\int d\rho\biggl\{\cos(kz -ft)\biggl\{\frac{\sin(4\phi)}{2}(A_{\times}-A_+)
\left({\bf e}_{\hat {\bf \rho}}\otimes 
{\bf e}_{\hat {\bf \phi}}+{\bf e}_{\hat {\bf \phi}}\otimes 
{\bf e}_{\hat {\bf \rho}}\right)+\label{e21} \\ && + 
\left(A_+\cos^2(2\phi)+A_{\times}\sin^2(2\phi)\right) 
 \biggl( {\bf e}_{\hat {\bf \rho}}\otimes 
{\bf e}_{\hat {\bf
\rho}}-{\bf e}_{\hat {\bf \phi}}\otimes 
{\bf e}_{\hat {\bf
\phi}}\biggr)\biggr\}-(M^P)^2(\phi)(f^P)_{\rho}^2(\rho)\biggr\}^{\frac{1}{2}}
\nonumber  \end{eqnarray} 
The expressions for $F^P(\rho,\phi)$, $F^P_{\rho}(\rho,\phi)$ and $h^P(\rho)$,  
given  by Eqs ({\ref{e19})-(\ref{e21}),  determine, as mentioned,  
the intrinsic geometry of the fringed trapped surface related to the  
plane GW's. \par
Now, as done for the optical  experiment from Section II and for 
the Brill case from
Section III, we find  here the orders and intervals 
related to the fringed  trapped surface geometry 
resulting from
the  plane GW's.  As
mentioned,  except for the superscript,   the
periodic $M^P(\phi)$ is identical to 
 $M^B(\phi)$ so relating the ordered bands of 
the fringed 
trapped surface to the same $\phi_m$ given by the same
expression $\phi_m=\phi_0 \pm m\pi$  
 (see the discussion before Eqs (\ref{e12})) 
one may write (compare with  Eqs (\ref{e12})) 
    \begin{eqnarray}  && x^P_{m}=
F^P(\rho,\phi_m)\cos(\phi_m) =f^P(\rho)M^P(\phi_m)\cos(\phi_m) \label{e22}  \\ 
&& y^P_{m}=
F^P(\rho,\phi_m)\sin(\phi_m)=f^P(\rho)M^P(\phi_m)\sin(\phi_m), \nonumber  
\end{eqnarray} 
where for the maximal bands one have either 
$-\pi  \le \phi \le -\frac{\pi}{2} $
 or 
$\frac{\pi}{2}  \le \phi \le \pi $ for which $M^P(\phi)=1$ (see Eq (\ref{e6})) and for  minimal
bands   $\phi$ should be from the range 
$-\frac{\pi}{2}  < \phi < \frac{\pi}{2}$ for which $M^P(\phi)=0$   (see Eq
(\ref{e6})).  
 Thus, denoting the interval between two neighbouring $m$ orders 
  in the fringed trapped surface
geometry  by $\Delta^P(\rho)$ and  using the relations 
$\cos(\phi_{(m+1)})=-\cos(\phi_m)$,  $\sin(\phi_{(m+1)})=-\sin(\phi_m)$ and 
$M^P(\phi_{(m+1)})+M^P(\phi_m)=2$  obtained from  Eq (\ref{e8})
one may write in correspondence with   Eq (\ref{e13})   
\begin{eqnarray}   && 
\Delta^P(\rho)= \sqrt{\left(x^P_{(m+1)}-x^P_{m}\right)^2+
\left(y^P_{(m+1)}-y^P_{m}\right)^2}= \nonumber \\ && = 
\biggl\{ \biggl(f^P(\rho)M^P(\phi_{m+1)})\cos(\phi_{(m+1)})-
f^P(\rho)M^P(\phi_m)\cos(\phi_m)\biggr)^2+ \nonumber \\ && +
\biggl(f^P(\rho)M^P(\phi_{m+1)})\sin(\phi_{(m+1)})-
f^P(\rho)M^P(\phi_m)\sin(\phi_m)\biggr)^2\biggr\}^{\frac{1}{2}}= \label{e23} \\ 
&& =
\biggl\{(f^P)^2(\rho)\biggl(M^P(\phi_{(m+1)})+
M^P(\phi_m)\biggr)^2\biggl(\cos^2(\phi_m)+\sin^2(\phi_m)\biggr)\biggr\}^{\frac{1}{2}}
=2f^P(\rho)  \nonumber \end{eqnarray}  
Using Eq (\ref{e19}) one may write the last equation as 
 \begin{eqnarray}   && 
\Delta^P(\rho)= 2f^P(\rho)=
\biggl\{ \biggl(\frac{4}{((M^P)^2(\phi)+(M^P)_{\phi}^2(\phi))}\biggr)
\cdot \label{e24} \\ && \cdot 
\rho^2\cos(kz -ft)  
\biggl[\frac{\sin(4\phi)}{2}(A_+-A_{\times}) 
 \cdot 
\left({\bf e}_{\hat {\bf \rho}}\otimes 
{\bf e}_{\hat {\bf \phi}}+{\bf e}_{\hat {\bf \phi}}\otimes 
{\bf e}_{\hat {\bf \rho}}\right) + \nonumber \\ && + 
\biggl(A_+\cos^2(2\phi)+  A_{\times}\sin^2(2\phi)\biggr)
 \biggl( {\bf e}_{\hat {\bf \phi}}\otimes 
{\bf e}_{\hat {\bf
\phi}}-{\bf e}_{\hat {\bf \rho}}\otimes 
{\bf e}_{\hat {\bf
\rho}}\biggr)\biggr] \biggr \}^{\frac{1}{2}}  
\nonumber \end{eqnarray}
  In Table 1 we have gathered
 in one place  the  expressions related to the DSE,  SSE and  
   the corresponding gravitational source-free Brill's and  plane 
   GW's.  The DSE case represents in this table the general $n$-slit experiment 
   where $n \geq 2$.   This table shows for all these 4 cases 
     the expressions related to the fringed and
   nonfringed situations.   The relevant
   expressions for the nonfringed trapped surfaces are derived in the
   appendices.    Note that for the nonfringed case each slit 
   of the DSE is
   treated as a separate SSE.   Also,   the interference results obtained 
   for the
   fringed case 
  from the, actually, single slit of the spurious double-slit screen 
   may, theoretically, be treated as if this single slit of length $b$ 
    is divided into two
   separate slits each of length $\frac{b}{2}$. Thus, the difraction of light
   from this single slit may be considered, for the fringed case, 
    as interference between light rays
   from the two halves of the slit as seen in the table.

   \markright{TABLE A}

     \begin{table}

\caption{\label{table1} In this  table we represent side by side  
the geometry of the optical
patterns of the DSE and SSE,  as seen on the surface of the photosensitive 
screen,  as well as  the
geometries of the trapped surfaces formed by the  source-free Brill's   and
 plane GW's. For all these  cases we show both the fringed
and nonfringed geometries and  also show   
the appropriate   changes  $\Delta$'s encountered when passing from the nonfringed
to the fringed situation.  The DSE case represents 
 the general $n$-slit experiment 
   where $n \geq 2$.   }
     \begin{center}
      \begin{tabular}{|l|l|l|l|l|} 
           Experimental &DSE  & SSE & Source-free  
	 & Source-free   \\
	 details && &Brill GW's& plane GW's   \\
        \hline \hline
\underline{The  nonfringed }  &  \underline{For each slit}  
     &     
       & $\scriptstyle{ x^B=f^B(\rho)\cos(\phi) } $     
 &$\scriptstyle{ x^P=f^P(\rho)\cos(\phi) }$ \\
 \underline{geometries}     &
   $\scriptstyle{  b\sin(\phi)_{{\cal B}}=m\lambda  }  $ 
&   $\scriptstyle{ b\sin(\phi)_{{\cal B}}=m\lambda }$    &  
$ \scriptstyle{ y^B=f^B(\rho)\sin(\phi)}$         & $\scriptstyle { 
y^P=f^P(\rho)\sin(\phi)}$ \\
  For both  DSE  and       & $\scriptstyle{ b\sin(\phi)_{{\cal
D}}=(m+\frac{1}{2})\lambda }$
    & $\scriptstyle{ b\sin(\phi)_{{\cal D}}= (m+\frac{1}{2})\lambda }$& 
 $\scriptstyle{z^B=h^B(\rho)}$  &$ \scriptstyle{z^P=h^P(\rho)}$ \\ 
 SSE the data 
    &$b=$ length of & $b=$ length of  
&  $\scriptstyle{ f^B(\rho)=\rho\psi^2} $&  
$ \scriptstyle{ f^P(\rho)=
\biggl\{ 
\rho^2\cos(kz -ft)\cdot } $  \\ 
through slits used.   & each slit.& slit.
&$ \scriptstyle{ h^B(\rho)=
\int d\rho\biggl(\psi^4e^{2Aq}- } $ & 
 $\scriptstyle{  \cdot \biggl[\frac{\sin(4\phi)}{2}(A_+-A_{\times}) 
 \cdot } $   \\
The optical patt-       & $b\sin(\phi)=$ path& $b\sin(\phi)=$ path&
$\scriptstyle{  - (f^B)_{\rho}^2(\rho)\biggr)}$& 
$ \scriptstyle{ \cdot \biggl({\bf e}_{\hat {\bf \rho}}\otimes 
{\bf e}_{\hat {\bf \phi}}+{\bf e}_{\hat {\bf \phi}}\otimes 
{\bf e}_{\hat {\bf \rho}}\biggr) +   } $ \\
-ern from both    &difference bet-  &difference bet- 
& 
&$\scriptstyle{ 
 \biggl(A_+\cos^2(2\phi)+ A_{\times}\sin^2(2\phi)\biggr) }$ \\
experiments are    & -ween rays from   &-ween rays from  & &
$ \scriptstyle{  \biggl( {\bf e}_{\hat {\bf \phi}}\otimes 
{\bf e}_{\hat {\bf
\phi}}-{\bf e}_{\hat {\bf \rho}}\otimes 
{\bf e}_{\hat {\bf
\rho}}\biggr)\biggr] \biggr \}^{\frac{1}{2}} }$ \\
of the diffract-    & ends of each slit.  & ends of slit.  &&\\
-ional nonfringed    & $m=0,\pm 1,\pm 2 \ldots$&$m=0,\pm 1,\pm 2,\ldots$
    &&$\scriptstyle{  h^P(\rho)=
\int d\rho\biggl\{\cos(kz -ft)} $\\ 
type.    & = orders of   
 &= orders of  &&$\scriptstyle{
  \biggl\{\frac{\sin(4\phi)}{2}(A_{\times}-A_+) }$ \\
The corresponding   & diffraction. &diffraction. &&$\scriptstyle{
\left({\bf e}_{\hat {\bf \rho}}\otimes 
{\bf e}_{\hat {\bf \phi}}+{\bf e}_{\hat {\bf \phi}}\otimes 
{\bf e}_{\hat {\bf \rho}}\right)+ } $\\ 
gravitational tra-    &  &&&$\scriptstyle{ 
\biggl(A_+\cos^2(2\phi)+A_{\times}\sin^2(2\phi)\biggr) } $ \\
-pped surfaces are    &  &    
      &&$\scriptstyle{ 
 \biggl( {\bf e}_{\hat {\bf \rho}}\otimes 
{\bf e}_{\hat {\bf
\rho}}-{\bf e}_{\hat {\bf \phi}}\otimes 
{\bf e}_{\hat {\bf
\phi}}\biggr)\biggr\}- } $ \\ 
also nonfringed. & &   
&&$\scriptstyle{ -(f^P)_{\rho}^2(\rho)\biggr\}^{\frac{1}{2}}}$\\
   &&&&\\
\hline
\underline{The fringed }  &  
$\scriptstyle{ d\sin(\phi)_{{\cal B}}=m\lambda}$
     & $\scriptstyle{\frac{b}{2}\sin(\phi)_{{\cal B}} =m\lambda}$
 & $ \scriptstyle{ x^B=f^B(\rho)M^B(\phi)\cos(\phi) }$&  
 $ \scriptstyle{ x^P=f^P(\rho)M^P(\phi)\cos(\phi) }$ \\
\underline{geometries}   & $\scriptstyle{ d\sin(\phi)_{{\cal D}}=(m+\frac{1}{2})\lambda}$   & 
 $\scriptstyle{ \frac{b}{2}\sin(\phi)_{{\cal D}}=(m+\frac{1}{2})\lambda}$  
  & $\scriptstyle{ y^B=f^B(\rho)M^B(\phi)\sin(\phi)}$    & 
$\scriptstyle{ y^P=f^P(\rho)M^P(\phi)\sin(\phi) }$\\
For both DSE       & 
  &  &$ \scriptstyle{ z^B(\rho)=h^B(\rho) }$ & 
$\scriptstyle{ z^P=h^P(\rho)  }$\\ 
and SSE the  data   &  & 
   &  
$\scriptstyle{ f^B(\rho)= }$ 
  &  
 $\scriptstyle{ f^P(\rho)=} $ \\ 
 about routes not    & $d=$interval bet- & $\frac{b}{2}=$ length of 
  &  
$\scriptstyle{ \frac{\rho\psi^2}{\sqrt{(M^B)^2(\phi)+(M^B)^2(\phi)_{\phi}}}}$ & 
$ \scriptstyle{ \biggl\{ \left(\frac{\rho^2\cos(kz -ft)}{(M^P)^2(\phi)+(M^P)^2(\phi)_{\phi}}\right)
 \cdot } $ \\
 used. The optical   & two slits.&half slit. &   $ \scriptstyle{ h^B(\rho)=
\int d\rho\biggl(\psi^4 e^{2Aq} - }$     & $ \scriptstyle{ \biggl[\frac{\sin(4\phi)}{2}(A_+-A_{\times}) 
 \cdot } $ \\
 patterns from &$d\sin(\phi)=$ path& $\frac{b}{2}\sin(\phi)=$ path&  
   $\scriptstyle{ - (f^B)_{\rho}^2(\rho)\biggr) }$   
   & $\scriptstyle{ \biggl({\bf e}_{\hat {\bf \rho}}\otimes 
{\bf e}_{\hat {\bf \phi}}+ {\bf e}_{\hat {\bf \phi}}\otimes 
{\bf e}_{\hat {\bf \rho}}\biggr) + } $ \\
 
\end{tabular} 
\end{center}
\end{table}

\begin{table}

     \begin{center}
      \begin{tabular}{|l|l|l|l|l|} 
          Experimental &DSE    & SSE & Source-free  
	 & Source-free  \\
	  details && &Brill  GW's&plane  GW's   \\
        \hline \hline

 both experi- &difference  betw- &difference betw- && 
 $\scriptstyle{ \biggl(A_+\cos^2(2\phi)+  A_{\times}\sin^2(2\phi)\biggr)\cdot }$ 
 \\
-ments  are of  &-een rays from  &-een rays from &           & 
  $\scriptstyle{  \cdot \biggl( {\bf e}_{\hat {\bf \phi}}\otimes 
{\bf e}_{\hat {\bf
\phi}}  - {\bf e}_{\hat {\bf \rho}}\otimes 
{\bf e}_{\hat {\bf
\rho}}\biggr)\biggr] \biggr \}^{\frac{1}{2}} }$  \\
the interference &two slits. &two halves of  && \\
fringed type. &&slit.&&  \\
 Corresponding   &$\scriptstyle{ \Delta(w) = 2g\tan(w) }$&$\scriptstyle{ \Delta(w) = 2g\tan(w) }$ 
   & $\scriptstyle{  \Delta^B(\rho)= 
\biggl\{\biggl(x^B_{(m+1)}-} $ &$\scriptstyle{ 
 h^P(\rho)=\int d\rho\biggl\{\cos(kz -ft) \cdot } $ \\ 
   gravitational    &&&  $\scriptstyle{-x^B_{m}\biggr)^2 +  
\biggl(y^B_{(m+1)}-  } $
  &   $\scriptstyle{ \cdot \biggl\{\frac{\sin(4\phi)}{2}
 (A_{\times}-A_+) \cdot }$   \\  
 trapped surfaces  &$m=0,\pm 1,$&$m=0,\pm 1,$& 
  $\scriptstyle{- y^B_{m}\biggr)^2 \biggr\}^{\frac{1}{2}}
 =\biggl\{ \biggl(f^B(\rho)\cdot }$  
  &$\scriptstyle{    \cdot \left({\bf e}_{\hat {\bf \rho}}\otimes 
{\bf e}_{\hat {\bf \phi}}+  {\bf e}_{\hat {\bf \phi}}\otimes 
{\bf e}_{\hat {\bf \rho}}\right)+ } $ \\  
are also fringed.     &$\pm 2 \ldots =$ orders  &$\pm 2 \ldots=$orders &  
$\scriptstyle{ \cdot M^B(\phi_{m+1)})
\cos(\phi_{(m+1)})-}$   
  &$ \scriptstyle{ + \biggl(A_+\cos^2(2\phi)+ A_{\times}\sin^2(2\phi)\biggr)
\cdot } $ \\ 
 The interference  &of interference. &of interference&  
$ \scriptstyle{-f^B(\rho)M^B(\phi_m) \cdot }$ 
  &$\scriptstyle{ \cdot
 \biggl( {\bf e}_{\hat {\bf \rho}}\otimes 
{\bf e}_{\hat {\bf
\rho}}-{\bf e}_{\hat {\bf \phi}}\otimes 
{\bf e}_{\hat {\bf 
\phi}}\biggr)\biggr\}- } $ \\ 
results for the &&between two &$\scriptstyle{ \cdot \cos(\phi_m)\biggr)^2
   +\biggl(f^B(\rho)\cdot }$ 
   &$ \scriptstyle{ -(M^P)^2(\phi)(f^P)_{\rho}^2(\rho)\biggr\}^{\frac{1}{2}} 
}$  
\\  
 SSE is theor- && halves of slit. & 
$ \scriptstyle{ \cdot M^B(\phi_{m+1)}) \sin(\phi_{(m+1)})
  -  } $ 
 &$\scriptstyle{ 
 \Delta^P(\rho)= \biggl\{\left(x^P_{(m+1)}-x^P_{m}\right)^2+ }$ \\
 -etically treated  &&   & 
 $\scriptstyle{ -f^B(\rho)
 M^B(\phi_m) \cdot }$ 
 &$\scriptstyle{ + 
\left(y^P_{(m+1)}-y^P_{m}\right)^2 \biggr\}^{\frac{1}{2}}= } $ \\ 
as if the single& &  &$\scriptstyle{
 \cdot \sin(\phi_m)\biggr)^2\biggr\}^{\frac{1}{2}}= 
 \biggl((f^B)^2(\rho) \cdot  } $ 
 &  $ \scriptstyle{ = 
\biggl\{ \biggl(f^P(\rho)M^P(\phi_{m+1)})\cdot }$ \\ 
 slit is divided    &&&$\scriptstyle{\cdot \biggl(M^B(\phi_{(m+1)})  
 +  } $      
& 
$ \scriptstyle{\cdot \cos(\phi_{(m+1)})- f^P(\rho)M^P(\phi_m) \cdot }$\\  
 into two halves.    &&&$\scriptstyle{+M^B(\phi_m)\biggr)^2\biggr)^{\frac{1}{2}}
  =2f^B(\rho) = }$ 
  &$ \scriptstyle{\cdot \cos(\phi_m)\biggr)^2  +\biggl(f^P(\rho) \cdot }$ \\ 
   Thus, the rays  &&& $\scriptstyle{ 
=\frac{2\rho\psi^2}{\sqrt{((M^B)^2(\phi)+(M^B)_{\phi}^2(\phi))}}}$ 
   &$\scriptstyle{\cdot M^P(\phi_{m+1)})\sin(\phi_{(m+1)}) -} $ \\
        from  these two        &&&  
 &$
 \scriptstyle{-f^P(\rho)M^P(\phi_m)\sin(\phi_m)\biggr)^2\biggr\}^{\frac{1}{2}}=
 }$\\ 
 halves may be 
  &&&$M^B(\phi), M^P(\phi)$    
 &$ \scriptstyle{=\biggl((f^P)^2(\rho)\biggl(M^P(\phi_{(m+1)})+}$ \\ 
thought as inter-  &&&given  by Eqs  &$\scriptstyle{  +M^P(\phi_m)\biggr)^2\biggr)^
{\frac{1}{2}}
= 2f^P(\rho)  = }$ \\ 
     fering with each    &&&(6) and (8).&$\scriptstyle{=\biggl\{ 
\biggl(\frac{4}{((M^P)^2(\phi)+(M^P)_{\phi}^2(\phi))}\biggr)
\cdot  }$ \\
 other.   &&&&$ \scriptstyle{ \cdot \rho^2\cos(kz -ft)  
\biggl[\frac{\sin(4\phi)}{2}(A_+-  }$ \\ 
  &&&&$\scriptstyle{-A_{\times}) \cdot \left({\bf e}_{\hat {\bf \rho}}\otimes 
{\bf e}_{\hat {\bf \phi}}+{\bf e}_{\hat {\bf \phi}}\otimes 
{\bf e}_{\hat {\bf \rho}}\right) + }$\\  
  &&&&$\scriptstyle{  +\biggl(A_+\cos^2(2\phi)+  A_{\times}\sin^2(2\phi)\biggr)
}$\\
 &&&&$\scriptstyle{
 \biggl( {\bf e}_{\hat {\bf \phi}}\otimes 
{\bf e}_{\hat {\bf
\phi}}-{\bf e}_{\hat {\bf \rho}}\otimes 
{\bf e}_{\hat {\bf
\rho}}\biggr)\biggr] \biggr\}^{\frac{1}{2}} } $ \\
&&&& \\
&&&& $M^P(\phi)$ given by \\
&&&& Eqs (6) and (8). \\
 \hline 
 \end{tabular} 
\end{center}
\end{table}

 \protect\section{Concluding Remarks}

 \markright{CONCLUDING REMARKS}

We  have represented and discussed a new variant of the DSE. 
 In this version thirty (30) mirrors   were prepared to serve as the relevant
screens  and a red laser pointer serves as a monochromatic light source. 
Twenty nine
(29)  mirrors were prepared as ordinary double-slit screens and the
remaining one 
 was, actually,  a single slit screen which  were
prepared so that it seemed as if it was a double one. After a large
number  of times  of repeating  the DSE upon these screens, where the
true nature of each of them,  whether it is the real  
double slit or
the faked one,   was not known during the experiments, one comes with the
unexpected result of obtaining interference pattern even when the screen was
later found to be the single-slit one. A photograph of this result is
shown in Figure 8.  That is, performing this experiment  under
the mentioned specific conditions has changed the form of the routes   
 through which the photons propagate between the two screens. Similar changes
 were shown, using any $n$-slit screen  ($n \geq 1$), 
   where it was established that if the data about the routes 
 through slits were used during the experiment one obtains  a number of
 diffraction patterns equal to the number of slits used (see Figures 2-4) 
   otherwise one obtains 
  the interference pattern of Figure 1.   That is, 
  if these data are used then most photons propagate in the forward directions 
  ($m=0$) for all $n$-slit screens ($n \geq 1$)   otherwise, 
   all
  directions are  
   equally traversed, even in the SSE,  and all  orders $m$ have the same 
   strength (see Figure 1). 
  \par 
  We have shown that  parallel gravitational
 situations,  in which the corresponding spacetime becomes fringed, 
 may, theoretically, be  related either  to the  
 source-free 
 plane GW's or  to the corresponding Brill's ones. For each of these  cases we have 
 used the method in \cite{eppley} and have calculated the appropriate fringed
 and nonfringed trapped surfaces by equating    
  their   metrics to the corresponding
 rotation metrics on the equator.  This is shown in the Brill's case by using  
  Eq (\ref{$A_{12}$}) of Appendix A for the nonfringed  trapped surface and  
   Eq (\ref{e10}) 
  of Section
  III for the fringed one. 
   In  the plane GW's case it is shown  by using  Eq (\ref{$B_{20}$}) of Appendix B
 for the nonfringed trapped surface and  Eq (\ref{e16}) of Section IV for 
 the fringed one. 
  This, of course,  does not mean that for the Brill case both  Eqations 
  of   (\ref{$A_{12}$}) 
   in Appendix A
  and Eq (\ref{e10})  are  valid at the same time or that for the plane GW case 
  both  Eqations of  
  (\ref{$B_{20}$})  in  Appendix B
  and Eq (\ref{e16})  are simultaneously valid since, as obviously
   seen, either pair of these equations are exclusive. One can not have both
  fringed and nonfringed trapped surfaces of the same kind existing side by side.   
 \par
As known,  the Bohr's complementarity principle \cite{merzbacher,schiff,bohr} explains why in
some experiments the particle nature of matter is demonstrated and in others 
its wave character  by stating that what determines the final actual
results of the experiment is what it is supposed to measure. This principle
assumes a thorough prior knowledge of all the constituents  of the experiments including,
of course, the true nature of the  screens activated in our optical examples. 
  We discuss here the case where 
 the observer does not know   
 the very  nature of these screens but think  that he is activating  
 (with 0.97 
probability)  a real  double-slit screen   
which is indeed realized  not only by obtaining an  interference
pattern over and over again  but also by checking these screens to find out 
that they are indeed double-slit. 
Thus, after obtaining the same result for hundreds of times the experiment
amounts, according to the observer,  to look for and find 
 this same optical pattern  which is, actually, what obtained even when the
 screen was later found to be single-slit. 
  Thus, one may argue that these results constitute a generalization 
 of the Bohr's 
complementarity principle in that they conform to what the observer 
{\it expects to
obtain  from the experiment}  even in case the activated apparatus is later 
found to be 
not  optimally suitable  for obtaining these
results. This is the meaning of writing  in the text, 
regarding the experiment described in Section II, 
that  knowing and, therefore, using
the mentioned  data  results in entirely different  consequences from those
obtained when these data are not used.   For example, 
as emphasized in the text regarding the case in which the spurious double-slit
screen was used,   the 
mere  knowledge during the experiment  of this  
 datum   entirely changes the character of the experiment including 
the  probability to obtain the corresponding diffraction pattern which is changed from 
the mentioned 0.033 to 1.    \par
  One may see this by
considering a changed version of the experiment   in which 
the element of not knowing the true nature of the chosen
screen  is absent so that the  
 the nature of the screen is checked immediately after randomely picking 
 it before
sending the laser ray through it. Thus, the experiment yields 
  one of two possible
results: (1) choosing a real double-slit mirror with a probability of
$\frac{29}{30}=0.97$  or 
(2) picking the faked  double-slit  mirror 
 with a probability of $\frac{1}{30}=0.033$.  In such case when (1) 
 is realized 
 the probability to find  interference pattern is unity and that for
 finding diffraction  one is zero. Similarly, one may see that if (2) is
 realized 
 the probability to find diffraction pattern is unity and that for
 finding  interference one is zero. Thus, denoting the wave functions for 
 choosing 
  the real and spurious double-slit screens by $\phi_1$ and $\phi_2$ 
  respectively 
  and 
 using the quantum mechanical
superposition principle \cite{merzbacher,schiff} one may write the 
corresponding wave function as
$ W=c_1\phi_1+c_2\phi_2,  $
where $c_1=\sqrt{\frac{29}{30}}$ and $c_2=\sqrt{\frac{1}{30}}$ are the
corresponding wave amplitudes. Thus, the
probability to find  interference pattern, even after repeating this
experiment thousands of times,  is always $(c_1\cdot 1)^2=\frac{29}{30}$ and  
that
for finding single-slit diffraction one is always 
$(c_2\cdot 1)^2=\frac{1}{30}$.  Note that  we have multiplied 
 $c_1$ and $c_2$ by unity
to emphasize that once the random choice of either (1) or (2) is done with the
respective probability amplitudes of $c_1$ and $c_2$ then the 
appropriate optical pattern follows in both cases with unity probability.  
\par  For the
experiment discussed in Section II, where the true nature of the chosen screen
is not known during the experiment, the two possible results are (compare with the 
former situation); (1) finding  
interference pattern with a probability of $\frac{29}{30}$  or (2) 
diffraction one with a probability of $\frac{1}{30}$. Thus, 
denoting the wave functions for finding  
    interference  and  diffraction patterns 
    by $\eta_1$ and $\eta_2$ respectively 
  and 
 using the quantum mechanical
superposition principle one may write the corresponding wave function as
$ W=c_1\eta_1+c_2\eta_2,  $ 
where $c_1=\sqrt{\frac{29}{30}}$ and $c_2=\sqrt{\frac{1}{30}}$ are the
corresponding wave amplitudes.  In 
such case one can not
revert the former discussion and say, for example,  that if (1) is realized then the
probability is unity to discover that the related screen is a  real 
double-slit and it is zero for  
finding a single-slit one. This is because the physical evolution always 
runs from 
first knowing the true character of the activated screen and only after
that to see the expected results obtained from  using this known screen. No one
guarantees that the reverted evolution of first seeing an  
interference  pattern and then ascertaining that the related screen is
double-slit is always followed. And indeed in Section II we have seen that 
under
specific conditions this evolution is not followed. \par
We note, in summary, that there is nothing special about the $n$-slit
experiments ($n \geq 1$)  which
causes their results to depend so critically upon using or not using during
these experiments the relevant nentioned data. That is, one may, logically,
suppose that similar results will also be obtained for other entirely different
experiments. For example, it is possible to design new versions of known
experiments in which some constituent elements are not known and, therefore, 
 not used
during these experiments. In such case one may, as for the previously discussed
$n$-slit experiments, expect to obtain results which are basically different 
from those
obtained when these elements are used.

 \markright{APPENDIX A}    
     
     \appendix 
     
\protect \section{APPENDIX A}
 
 \bigskip \bigskip    
     
\protect \subsection{ The canonical formalism of the general relativity 
and the Brill GW's}

\bigskip 

We represent here, for completness, a short review of the ADM formalism
\cite{mtw} and its
adaptation to the source-free Brill wave \cite{mtw,brill1,gentle,eppley}. In the ADM
canonical formulation of general relativity one starts from the (3+1)-dimensional split 
of space-time which is expressed by the corresponding metric
tensor \cite{mtw,adm}

\begin{equation}  ^{(4)}g_{\alpha\beta}=
\left(  \begin{array}{cc} ^{(4)}g_{00} &  ^{(4)}g_{0j} \\ 
^{(4)}g_{i0} & ^{(4)}g_{ij}  \end{array} \right)   =  \left( 
\begin{array}{cc}   N_kN^k-N^2 & N_j \\ N_i & g_{ij}  \end{array} 
\right),   \tag{$A_1$}  \label{$A_1$}  \end{equation}
where the  spacelike three-dimensional hypersurfaces at constant times 
are represented by   the metric tensor $g_{ij}$. The shift vector is denoted by 
$N_i$ and the lapse function by $N$. Denoting the covariant derivative by $|$ 
one may write the action \cite{mtw}
\begin{equation} I=\frac{1}{16\pi}\int  \biggl( \pi^{ij}\frac{\partial
g_{ij}}{\partial t}-N{\cal H}(\pi^{ij},g_{ij})-N_i{\cal H}^i(\pi^{ij},g_{ij}) 
\biggr) d^4x, \tag{$A_2$}   \label{$A_2$} \end{equation}
where the superhamiltonian ${\cal H}$ and supermomentum ${\cal H}^i$ are
\cite{mtw}
\begin{align}   {\cal
H}(\pi^{ij},g_{ij})&=g^{-\frac{1}{2}}(T\!r(\pi^2)-\frac{1}{2}(T\!r
({\bf \pi}))^2)-g^{\frac{1}{2}}R \tag{$A_3$}  \label{$A_3$}  \\ 
 {\cal H}^i(\pi^{ij},g_{ij})&=
-2\pi^{ik}_{|k}  \nonumber \end{align} 
 Upon extremization of the action $I$ with respect to $g_{ij}$ and
$\pi^{ij}$  one may write the following vacuum field equations \cite{mtw} 
\begin{equation}  \frac{\partial g_{ij}}{\partial t}
=2Ng^{-\frac{1}{2}}\biggl( \pi_{ij}-\frac{1}{2}g_{ij}T\!r 
({\bf \pi}) \biggr)+N_{i|j}+N_{j|i} \tag{$A_4$}  \label{$A_4$} 
\end{equation} 
\begin{align}    &\frac{\partial \pi^{ij}}{\partial
t}=-N\biggl(R^{ij}-\frac{1}{2}g^{ij}R \biggr)+
\frac{1}{2}Ng^{-\frac{1}{2}}g^{ij}\biggl( T\!r({\bf 
\pi}^2)-
\frac{1}{2}(T\!r ({\bf \pi}))^2\biggr) -  \nonumber \\   
 -& 2Ng^{-\frac{1}{2}}\biggl(\pi^{im}\pi^j_m-\frac{1}
{2}\pi^{ij}T\!r
({\bf \pi})\biggr) + g^{ij}\biggl( N^{|ij}-g^{ij}N^{|m}_{|m} \biggr)
-(\pi^{ij}N^m)_{|m} -   \tag{$A_5$}   \label{$A_5$} \\  
  - &N^i_{|m}\pi^{mj}-N^j_{|m}\pi^{mi}    \nonumber
\end{align} 
The appropriate initial-value equations in this ADM formalism are obtained 
upon
extremization of the action $I$  from Eq (\ref{$A_2$}) with respect 
to the lapse $N$ and shift $N_i$. 
Thus, taking into account that  the extrinsic curvature tensor is given by
\cite{mtw} 
$K_{ij}=\frac{1}{2N}(N_{i|j}+N_{j|i}-\frac{\partial g_{ij}}{\partial t})$  
one may write, for the vacuum case, upon
extremization with respect to the lapse $N$   the initial-value condition  
  \cite{mtw}
\begin{equation}  ^{(3)}R+(T\!r({\bf K}))^2-T\!r({\bf K}^2)=0 \tag{$A_6$} 
\label{$A_6$} 
\end{equation} 
And upon extremization with respect to the
shift $N_i$ one obtains the three initial-value conditions   \cite{mtw} 
\begin{equation}   (K^k_i)_{|k}-\delta^k_i(T\!r({\bf K}))_{|k}=0 \tag{$A_7$} 
\label{$A_7$} 
\end{equation}    
For easing the following discussion we assume that the relevant space-time is
characterized by time and axial symmetries and no rotation. The time symmetry
property ensures  \cite{mtw,brill1,gentle,eppley} the existence of a spacelike hypersurface 
${\bf \Sigma}$ in
which the extrinsic curvature ${\bf K}_{ij}$ vanishes at all its points. 
In such
case the three momentum initial conditions from Eq (\ref{$A_7$}) are trivially
satisfied and the fourth Hamiltonian initial condition from Eq (\ref{$A_6$}) 
reduces, for the vacuum case, to $^{(3)}R=0$. Thus, as in \cite{brill1}, we take
the following conformal basic metric on the initial spacelike hypersurface 
\begin{equation}  ds^2_{conformal}=e^{2Aq(\rho,z)}(dz^2+d\rho^2)+
\rho^2d\phi^2   \tag{$A_8$}  \label{$A_8$}  \end{equation}  
From this metric one obtains the following components for the metric tensor
\cite{brill1,eppley}
$g_{ij}$ \begin{equation}     g_{\rho\rho}=g_{zz}=e^{2Aq(\rho,z)}, \ \ \
g_{\phi\phi}=\rho^2, \ \ \ g_{\rho z}=g_{\rho\phi}=g_{z\phi}=0,   
\tag{$A_9$}   \label{$A_9$}      \end{equation}    
where $g_{\rho\phi}=g_{z\phi}=0$ follow from the no-rotation assumption. The
axial symmetry property ensures that on the $z$ axis, where $\rho=0$, the
function $q$ should vanish. The hamiltonian initial condition is solved, as in
\cite{brill1,eppley}, by assuming the following conformal map 
\begin{equation}    ds^2_{physical}=\psi^4ds^2_{conformal}=
\psi^4\Bigl(e^{2Aq(\rho,z)}(dz^2+d\rho^2)+
\rho^2d\phi^2\Bigr),  \tag{$A_{10}$}     \label{$A_{10}$}
\end{equation}    
 where $ \psi$ is the conformal factor.  The 
 embedded surface  is obtained by assuming the metric of the equator to be equal
 to 
 that of a surface of rotation in Euclidean space \cite{eppley} defined as  
  \begin{equation} x^B=f^B(\rho)\cos(\phi), \ \ y^B=f^B(\rho)\sin(\phi), 
  \ \ z^B=h^B(\rho),  
 \tag{$A_{11}$}  \label{$A_{11}$} \end{equation} 
 where the superscript $B$ denotes that we consider the Brill  GW's. 
 Noting that on the equator $g_{zz}=0$ one may perform the  equating process 
 as \cite{eppley}
 \begin{align}    
 (&ds^B)^2=(dx^B)^2+(dy^B)^2+(dz^B)^2=
 ((f^B)_{\rho}^2(\rho)+(h^B)_{\rho}^2(\rho))d\rho^2+ 
 \tag{$A_{12}$}  \label{$A_{12}$} \\ 
+ &(f^B)^2(\rho)d\phi^2= 
 g^B_{{\hat {\bf \rho}}{\hat{\bf \rho}}}d^2\rho +
g^B_{{\hat {\bf \phi}}{\hat{\bf \phi}}}d^2\phi =
 \psi^4\biggl(e^{2Aq(\rho,z)}d\rho^2+
\rho^2d\phi^2\biggr) \nonumber \end{align}
Thus, the surface on the equator which is characterized by the same intrinsic
geometry as that of the mentioned generating Brill GW is  \cite{eppley} 
\begin{align}
 f^B(\rho)&=\rho\psi^2 \nonumber \\
 f^B_{\rho}(\rho)&=\psi^2+2\rho\psi\psi_{\rho}   
  \tag{$A_{13}$}  \label{$A_{13}$} \\
 h^B(\rho)&=\int d\rho\biggl(\psi^4e^{2Aq}-(f^B)^2_{\rho}\biggr) \nonumber 
\end{align} 

Schematic representations of
such three surfaces are shown in Figure 11 for the amplitudes $A \approx 2$, 
$A \approx 5$, $A \approx 15$. The circular-form surface at the
bottom corresponds to the smaller amplitude $A \approx 2$ whereas the upper
pinched-off surface corresponds to $A \approx 15$. That is, as the amplitude $A$
of the wave increases the surface deviates from the circular form and tends to
be closed  upon itself (pinched-off). Note, however, that these embeddings do
not determine the exact amplitude and shape of the developed apparent throats.
For this one have to solve the trapped surface equation (see, for example, Eq
(27) in \cite{eppley}, see also \cite{brill2} (which represents another 
embedding
method)). This is generally done by using
numerical analysis \cite{alcubierre} through which one may develop and follow 
the Brill initial
data across a grid framework.

\begin{figure}
\centerline{
\begin{turn}{-180}
\includegraphics[width=12cm]{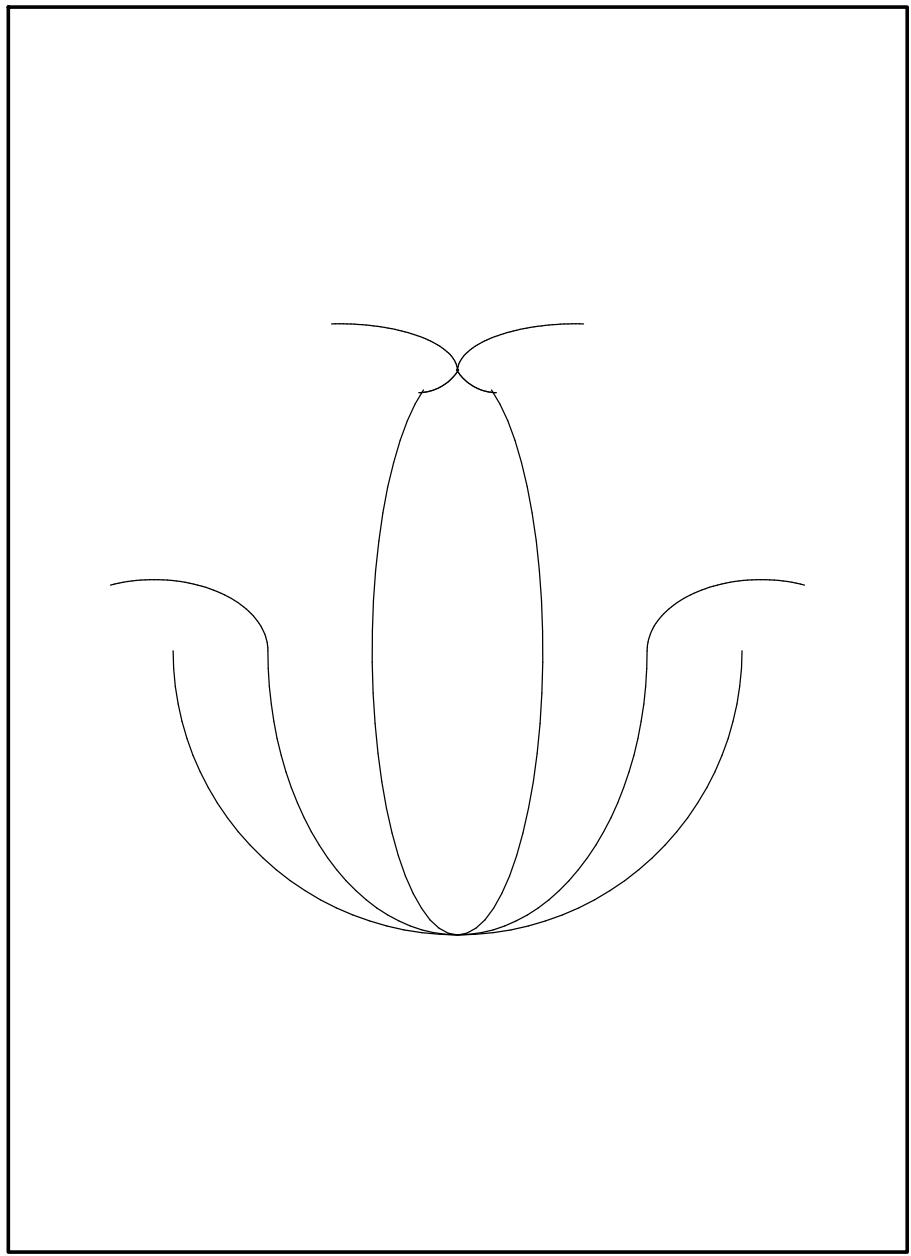}
\end{turn}}

     \caption{A schematic representation of the 
      embedded trapped surface shown for the three values of the
     amplitude: $A \approx 2$,  $A \approx 5$  and  $A \approx 15$.  
     Note that the circular form of the
     bottom surface corresponds to the smallest amplitude of $A \approx 2$ 
     and that as this amplitude grows the 
     surface tends to
     pinch off and to close on itself.}
     \end{figure}

\markright{APPENDIX B}

\protect \section{APPENDIX B}

\protect \subsection{ The  linearized plane  GW's and geodesics mechanics}

In this Appendix we review the theory of plane gravitational wave as represented
in \cite{bar1}. For that purpose we first introduce the basic theory of 
geodesics mechanics in the
presence of GW's  as outlined in \cite{mtw}. That is, we calculate the change in
location of a test point (TP) moving along its geodesic route relative to
another TP due to a passing GW. The two TP's, as well as their specific
geodesics, are denoted by ${\cal A}$ and ${\cal B}$ and the interval between
them is denoted by the vector ${\bf n}$.  
 The proper reference frame of ${\cal A}$ is chosen as the
appropriate coordinate system. That is, the spatial origin $x^j=0$ is attached to the
world line of ${\cal A}$ and the coordinate time $x^0$ is  
${\cal A}$'s proper
time so that  $x^0=\tau$ on the world line $x^j=0$ (see Chapter 35 in 
\cite{mtw}).  This system is assumed to be nonrotating frame as that 
 obtained by
attaching the orthonormal spatial axes to gyroscopes  \cite{mtw}.  Thus,
it constitutes a local Lorentz frame \cite{mtw,bergmann}
along the whole world line of ${\cal A}$ and not only at one event  on it. 
As mentioned, we use  the linearized theory of gravitation \cite{mtw} and, therefore, 
 the
metric tensor is \cite{mtw}
\begin{equation}   g_{\mu \nu}=\eta_{\mu \nu}+h_{\mu \nu}+O(h_{\mu \nu})^2, 
 \tag{$B_1$}   \label{$B_1$}     \end{equation} 
  where $\eta_{\mu \nu}$ is the  Lorentz metric tensor \cite{mtw,bergmann} 
  and   $h_{\mu \nu}$ 
   is a slight perturbation  of it. 
    The relevant metric is \cite{mtw}
\begin{equation}   
ds^2=-(dx^0)^2+\delta_{jk}dx^jdx^k+O(|x^j|^2)dx^{\alpha}dx^{\beta} 
\tag{$B_2$}   \label{$B_2$}
\end{equation}   
The small
perturbation $h_{\mu \nu}$ from Eq (\ref{$B_1$})  is  identified,  
as done in \cite{mtw,thorne},   with the passing GW.    Use  is made of the transverse-traceless (TT) 
gauge \cite{mtw,thorne}  which have the following properties; (1) all  components vanish except the
spatial ones, i.e.,
$h^{TT}_{\mu 0}=0$, (2) these components are
divergence-free i.e., $h^{TT}_{kj,j}=0$, and (3) they are trace-free i.e., 
$h^{TT}_{kk}=0$. Also, as emphasized in \cite{mtw},  only 
pure GW's, of the kind discussed here, can be reduced to TT gauge.  
 As noted, the GW is identified with 
$h_{jk}^{TT}$ and, therefore,  have the same characteristics  \cite{mtw}. \par
The world lines  ${\cal A}$ and ${\cal B}$  represent geodesics and so 
TP's fall freely along them. Thus,   
assuming  the basis $e_{\beta}$ changes
arbitrarily but smoothly from point to point    
one may write the velocity of
the TP  ${\cal B}$ relative to  ${\cal A}$ as \cite{mtw}
$ \nabla_{\bf u}{\bf n}=(n^{\beta};_{\gamma} u^{\gamma})e_{\beta}$,
   where ${\bf n}$  is the vector from ${\cal A}$ to ${\cal B}$, 
 ${\bf u}$ is the tangent vector to the geodesic ${\cal B}$ i.e., 
   ${\bf u}=\frac{\partial {\cal B}({\bf n},\tau)}{\partial \tau}$ and   
$n^{\beta};_{\gamma}$ is the covariant derivative of $n^{\beta}$   
\cite{mtw,bergmann} i.e., 
$
n^{\beta};_{\gamma}=\frac{dn^{\beta}}{dx^{\gamma}}+\Gamma^{\beta}_{\mu
\gamma}n^{\mu}  $ where  
  $\Gamma^{\beta}_{\mu \gamma}$ is  \cite{mtw,bergmann} 
$ \Gamma^{\beta}_{\mu \gamma}=g^{\nu \beta}\Gamma_{\nu \mu
\gamma}=\frac{1}{2}g^{\nu \beta}(g_{\nu \mu,\gamma}+g_{\nu \gamma,\mu}-g_{\mu
\gamma,\nu})$. 
 The expression inside the circular parentheses 
 $n^{\beta};_{\gamma} u^{\gamma}$ represents 
 the
components of $\nabla_{\bf u}{\bf n}$  \cite{mtw}
i.e., $n^{\beta};_{\gamma} u^{\gamma}= 
\frac{Dn^{\beta}}{d\tau} 
=\frac{dn^{\beta}}{d\gamma}+\Gamma^{\beta}_{\mu
\gamma}n^{\mu}\frac{dx^{\gamma}}{d\tau}$ so 
  the  expression for the  acceleration of the TP ${\cal B}$ 
relative to  
${\cal A}$  \cite{mtw} 
$ \nabla_{\bf u}(\nabla_{\bf u}{\bf n})=-R,$ 
 may be written  componently    as \cite{mtw} 
\begin{equation}   \frac{D^2n^{\alpha}}{d\tau^2}=
-R^{\alpha}_{\beta \gamma \delta}u^{\beta}u^{\delta}n^{\gamma}, 
\tag{$B_3$}   \label{$B_3$} 
\end{equation}  
where $R$ is the Riemann curvature tensor whose components are written  as 
\cite{mtw,bergmann}
$ R^{\alpha}_{\beta \gamma \delta}=\frac{\partial
\Gamma^{\alpha}_{\beta \delta}}{\partial x^{\gamma}}-\frac{\partial
\Gamma^{\alpha}_{\beta \gamma}}{\partial x^{\delta}}+\Gamma^{\alpha}_{\mu
\gamma}\Gamma^{\mu}_{\beta \delta}-\Gamma^{\alpha}_{\mu
\delta}\Gamma^{\mu}_{\beta \gamma}$.
Now, remembering that  $x^0=\tau$ on the world
line $x^j=0$ of ${\cal A}$  one may write  (\ref{$B_3$}) as
 \begin{equation}   \frac{D^2n^j}{d\tau^2}=
-R^j_{0k0}n^k=-R_{j0k0}n^k  
\tag{$B_4$}   \label{$B_4$}
\end{equation} 
 Note \cite{mtw} that, to first
order in the metric perturbation $h^{TT}_{jk}$,  the transverse trace-free  
(TT) coordinate system 
 may move  \cite{mtw} with the proper
reference frame of ${\cal A}$.  
 That is, to this order in $h^{TT}_{jk}$,  the time $t$ in the system TT 
  may be identified  \cite{mtw} with the proper time $\tau$ of  
 ${\cal A}$
so that \cite{mtw}  $R^{TT}_{j0k0}=R_{j0k0}$ where  
 $R_{j0k0}$  is
calculated in ${\cal A}$'s proper reference frame  and  $R^{TT}_{j0k0}$, which 
  is
calculated in the $TT$  system,  were shown 
(see Eq (35.10) in \cite{mtw}) to  assume the  simple
form 
 of $ R_{j0k0} = -\frac{1}{2}h^{TT}_{jk,00} $.  
 Note  that since the $TT$  system and  the proper
 reference frame of  ${\cal A}$  move together they are both denoted \cite{mtw} 
 by 
 the same  indices $(0,k,j)$ 
 with no need  to use primed and unprimed indices.    
Also,  since the origin is situated along ${\cal A}$'s geodesic the 
components of the
separating vector ${\bf n}$ are no other than  the coordinates of ${\cal B}$.  
That is, 
writing the coordinates of ${\cal A}$ and ${\cal B}$ as  $x^j_{{\cal A}}$ and 
$x^j_{{\cal B}}$ 
 one obtains  
$n^j=x^j_{{\cal B}}-x^j_{{\cal A}}=x^j_{{\cal B}}-0=x^j_{{\cal B}}$. Also, 
 at $x^j=0$  we have $\Gamma^{\mu}_{\alpha \beta}=0$ for all $x^0$ 
 so that $\frac{d\Gamma^{\mu}_{\alpha \beta}}{d\tau}=0$  and  the covariant
derivative $\frac{D^2n^j}{d\tau^2}$ reduces \cite{mtw} to  ordinary derivative. 
   That is,  Eq (\ref{$B_4$}) becomes  \begin{equation}  \frac{d^2x^j_B}{d\tau^2}=
-R_{j0k0}x^k_{{\cal B}} =\frac{1}{2}(\frac{\partial ^2h^{TT}_{jk}}{\partial
t^2})x^k_{{\cal B}} 
\tag{$B_5$}   \label{$B_5$}
\end{equation} 
As  initial condition we assume the TP's ${\cal A}$ and 
${\cal B}$ to be  at
rest before the GW arrives.  That is, $x^j_{{\cal B}}=x^j_{{\cal B}(0)}$ when 
$h^{TT}_{jk}=0$ so that 
 the solution of Eq (\ref{$B_5$}) is
\begin{equation}    x^j_{{\cal B}}(\tau)=x_{{\cal B}(0)}^k(\delta_{jk}+\frac{1}{2}h^{TT}_{jk})_{at
{\cal A}}, \tag{$B_6$}   \label{$B_6$}  \end{equation} 
which is  the new location of ${\cal B}$   as seen 
in the proper reference frame of ${\cal A}$.  The $h^{TT}_{jk}$ 
   represents the  
passing GW  which is supposed here to be  a plane 
 wave advancing in the $\hat{{\bf
n}}$
direction (not  the same as the separation vector ${\bf n}$) 
where the TP's ${\cal A}$ and ${\cal B}$ and their relevant  geodesics 
 lie in the plane perpendicular to
$\hat{{\bf n}}$.  We denote the two perpendicular directions 
to $\hat{{\bf n}}$ by 
${\bf e}_{\hat{{\bf n}}_1}$ and ${\bf e}_{\hat{{\bf n}}_2}$ and note 
\cite{mtw} that 
this GW have the following unit
 linear-polarization tensors  
\begin{equation} {\bf e}_{+_{\hat{{\bf n}}_1\hat{{\bf n}}_1}}= 
 {\bf e}_{\hat{{\bf n}}_1} \otimes {\bf e}_{\hat{{\bf n}}_1}- {\bf e}_{\hat{{\bf
 n}}_2} \otimes {\bf e}_{\hat{{\bf n}}_2}= -\biggl({\bf e}_{\hat{{\bf n}}_2} \otimes
 {\bf e}_{\hat{{\bf n}}_2}- {\bf e}_{\hat{{\bf
 n}}_1} \otimes {\bf e}_{\hat{{\bf n}}_1}\biggr)=-{\bf e}_{+_{\hat{{\bf n}}_2\hat{{\bf
 n}}_2}}  \tag{$B_7$}  \label{$B_7$}   \end{equation}
 \begin{equation}  {\bf e}_{\times_{\hat{{\bf n}}_1\hat{{\bf n}}_2}}= 
 {\bf e}_{\hat{{\bf n}}_1} \otimes {\bf e}_{\hat{{\bf n}}_2}+ {\bf e}_{\hat{{\bf
 n}}_2} \otimes {\bf e}_{\hat{{\bf n}}_1}= 
\biggl( {\bf e}_{\hat{{\bf n}}_2} \otimes {\bf e}_{\hat{{\bf n}}_1}+ {\bf e}_{\hat{{\bf
 n}}_1} \otimes {\bf e}_{\hat{{\bf n}}_2}\biggr)= {\bf e}_{\times_{\hat{{\bf
 n}}_2\hat{{\bf n}}_1}},  \tag{$B_8$}  \label{$B_8$} \end{equation} 
   where $\otimes$  is \cite{spiegel2,synge} the tensor product. Thus, considering the mentioned 
    $TT$'s 
gauge  constraints 
 $h^{TT}_{\mu 0}=0$,  $h^{TT}_{ij,j}=0$   and  $h^{TT}_{kk}=0$  
 one may realize  that  for the GW propagating
 in the  $\hat{{\bf n}}$ direction, the only nonzero components  
  are \cite{mtw} 
\begin{align}   &  h^{TT}_{+_{{\hat{{\bf n}}_1\hat{{\bf n}}_1}}} =\Re 
\biggl( A_+{\bf e}_{+_{\hat{{\bf n}}_1\hat{{\bf n}}_1}}e^{-ift}
e^{ik{\bf r}\hat{{\bf n}}} \biggr) = 
 A_+{\bf e}_{+_{\hat{{\bf n}}_1\hat{{\bf
n}}_1}} \cdot  
\cos\biggl(k\Big({\bf r}_1\cos(\alpha)+\nonumber \\  + &
{\bf r}_2\cos(\beta)+ 
{\bf r}_3\cos(\eta)\Bigr)-  
 ft\biggr) = -h^{TT}_{+_{{\hat{{\bf n}}_2\hat{{\bf n}}_2}}}=-\Re 
\biggl( A_+{\bf e}_{+_{\hat{{\bf n}}_2\hat{{\bf n}}_2}}e^{-ift}
e^{ik{\bf r}\hat{{\bf n}}} \biggr) =  \tag{$B_9$}  \label{$B_9$}   \\ 
 =& - A_+{\bf e}_{+_{\hat{{\bf n}}_2\hat{{\bf
n}}_2}} \cdot  
\cos\biggl(k\Big({\bf r}_1\cos(\alpha)+{\bf r}_2\cos(\beta)+{\bf r}_3\cos(\eta)\Bigr)-ft\biggr)  
\nonumber \end{align}  
\begin{align} &  h^{TT}_{\times_{{\hat{{\bf n}}_1\hat{{\bf n}}_2}}}=\Re 
\biggl( A_{\times}{\bf e}_{\times_{\hat{{\bf n}}_1\hat{{\bf n}}_2}}e^{-ift}
e^{ik{\bf r}\hat{{\bf n}}} \biggr) = 
 A_{\times}{\bf e}_{\times_{\hat{{\bf n}}_1\hat{{\bf
n}}_2}} \cdot  
\cos\biggl(k\Bigl({\bf r}_1\cos(\alpha)+ \nonumber \\ +& 
{\bf r}_2\cos(\beta)+  
+ {\bf r}_3\cos(\eta)\Bigr) 
- ft\biggr) 
= h^{TT}_{\times_{{\hat{{\bf n}}_2\hat{{\bf n}}_1}}}=\Re 
\biggl( A_{\times}{\bf e}_{\times_{\hat{{\bf n}}_2\hat{{\bf n}}_1}}e^{-ift}
e^{ik{\bf r}\hat{{\bf n}}} \biggr) =  \tag{$B_{10}$}  \label{$B_{10}$} \\ 
  = &
 A_{\times}{\bf e}_{\times_{\hat{{\bf n}}_2\hat{{\bf
n}}_1}} \cdot  
\cos\biggl(k\Bigl({\bf r}_1\cos(\alpha)+{\bf r}_2\cos(\beta)+{\bf r}_3\cos(\eta)\Bigr)-ft\biggr) 
\nonumber \end{align}  
where $\Re$ denotes the real parts  of the expressions which follow and 
 ${\bf r}$
is the position vector of a point in space.     By    $A_+$
and $A_{\times}$  we denote the amplitudes which are   respectively  
 related to  the  two 
 modes of
polarization ${\bf e}_{+_{\hat{{\bf n}}_1\hat{{\bf n}}_1}}$ and 
${\bf e}_{\times_{\hat{{\bf n}}_1\hat{{\bf n}}_2}}$.  By 
$f$ we denote the time frequency,  $k$ is
$\frac{2\pi}{\lambda}$,  and 
$\cos(\alpha),\ \cos(\beta), \ \cos(\eta)$ are the
direction cosines of  $\hat{{\bf n}}$.  Thus,  the general perturbation
$h^{TT}_{jk}$ resulting from the passing GW may be written as  \cite{bar1,bar2}
\begin{align}  &  h^{TT}_{jk}= h^{TT}_{+_{jk}}+
h^{TT}_{\times_{jk}}=\Re 
\biggl( (A_+{\bf e}_{+_{jk}}+
A_{\times}{\bf e}_{\times_{jk}})e^{-ift}e^{ik{\bf
r}\hat{{\bf n}}}\biggr) =    \tag{$B_{11}$}  \label{$B_{11}$}  \\ 
  =& (A_+{\bf e}_{+_{jk}}+
A_{\times}{\bf e}_{\times_{jk}})
\cos\biggl(k\Bigl({\bf r}_1\cos(\alpha)+{\bf r}_2\cos(\beta)+{\bf r}_3\cos(\eta)\Bigr)-ft\biggr) 
\nonumber \end{align} 
A better understanding of the situation follows when one considers \cite{mtw} an ensemble
of TP's. That is, assuming a large number     
 of TP's ${\cal B}$ which form a 
circular (elliptic) ring
around the TP ${\cal A}$ in the center one may realize that    
the  passing GW with either
${\bf e}_{+{\hat{{\bf n}}_1\hat{{\bf n}}_1}}$  or 
${\bf e}_{\times_{\hat{{\bf n}}_1\hat{{\bf n}}_2}}$  polarization 
periodically changes the former array into an  elliptic 
(circular) one
 as shown in Figure 12.

\begin{figure}
\centerline{
\begin{turn}{-90}
\includegraphics[width=12cm]{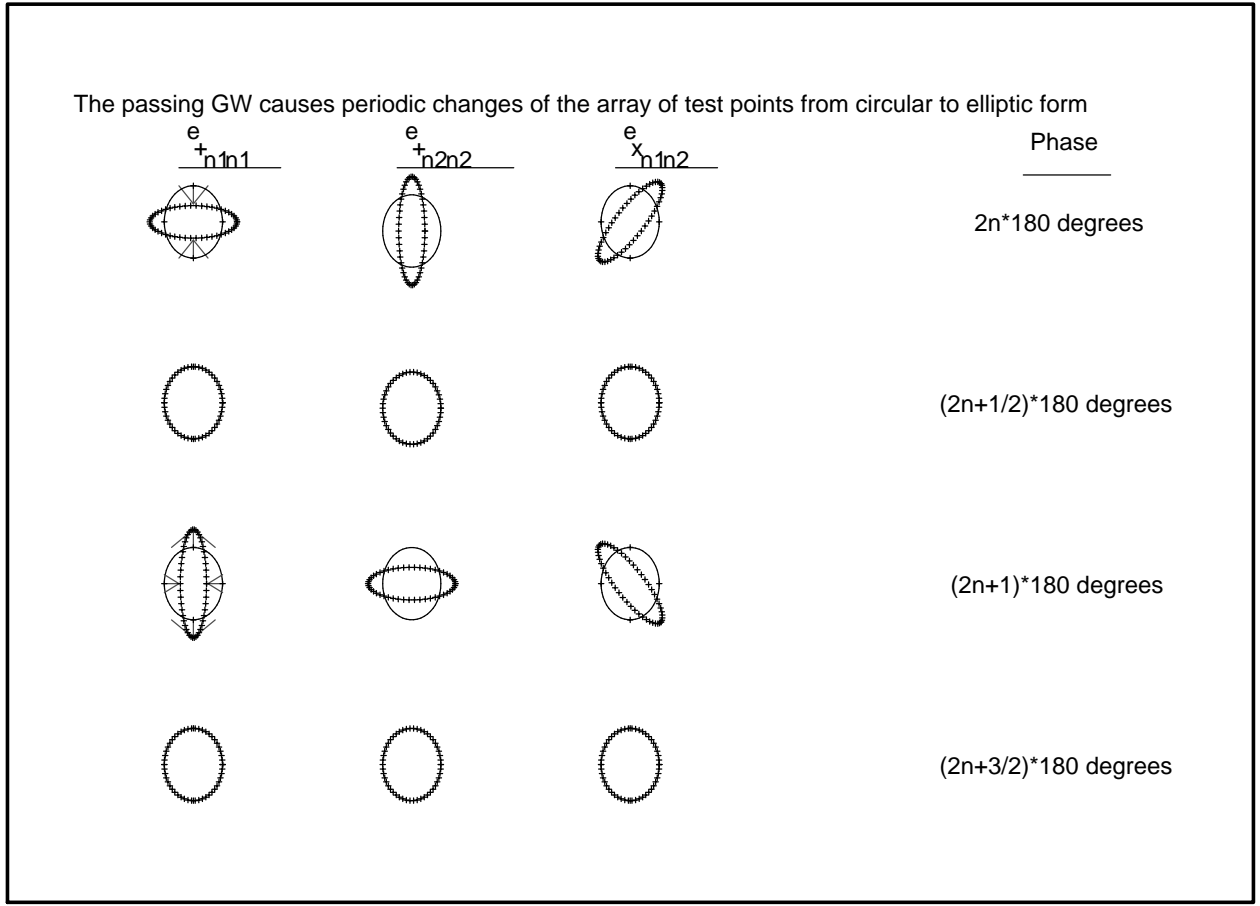}
\end{turn}}

     \caption{A schematic representation of the influence of a passing plane GW
upon a circular (elliptic) array of test particles which periodically changes
its form to elliptic (circular) array. }
     \end{figure}

Substituting 
from Eq (\ref{$B_{11}$})  into  Eq
(\ref{$B_6$})      
 one obtains the components along ${\bf n}_1$  and  ${\bf n}_2$ of the new
 location of the TP ${\cal B}$ as calculated in the proper reference frame of 
${\cal A}$. That is,  for $j={\bf n}_1$ one obtains \cite{bar1}
   \begin{align}     &&  x^{{\hat {\bf n_1}}}_{{\cal B}}=
   \Biggl\{x_{{\cal B}(0)}^{{\hat {\bf n_1}}}+\frac{1}{2}\biggl( 
   A_+{\bf e}_{+_{\hat{{\bf n}}_1\hat{{\bf n}}_1}}x_{{\cal B}(0)}^{{\hat {\bf n_1}}}+
A_{\times}{\bf e}_{\times_{\hat{{\bf n}}_1\hat{{\bf n}}_2}}x_{{\cal B}(0)}^{{\hat {\bf
n_2}}}\biggr) \cdot  \tag{$B_{12}$}  \label{$B_{12}$}  \\    
&& \cdot 
\cos\biggl( k\biggl(x\cos(\alpha)+y\cos(\beta)+z\cos(\eta)\biggr)-ft\biggr)
\Biggr\}_{at {\cal A}}  
\nonumber  \end{align}    
And for $j={\bf n}_2$  one obtains \cite{bar1}
   \begin{align}    &&  x^{{\hat {\bf n_2}}}_{{\cal B}}=
   \Biggl\{x_{{\cal B}(0)}^{{\hat {\bf n_2}}}+\frac{1}{2}\biggl( 
   A_{\times}{\bf e}_{\times_{\hat{{\bf n}}_2\hat{{\bf n}}_1}}x_{{\cal B}(0)}^{{\hat {\bf n_1}}}+
A_+{\bf e}_{+_{\hat{{\bf n}}_2\hat{{\bf n}}_2}}x_{{\cal B}(0)}^{{\hat {\bf
n_2}}} \biggr) \cdot \tag{$B_{13}$}  \label{$B_{13}$}   \\ 
&& \cdot 
\cos\biggl( k\Bigl(x\cos(\alpha)+y\cos(\beta)+z\cos(\eta)\Bigr)-ft\biggr)
\Biggr\}_{at {\cal A}}    \nonumber 
\end{align}

 \protect  \subsection{\large The nonfringed trapped surface resulting from 
 plane GW's}

As shown in \cite{bar1} the comparison between the electromagnetic (EM) 
theory and the
linearized general relativity enables one to use theoretical methods similar to
those used in the EM theory for assuming interference and holographic properties
for GW's also. Similar comparison  between these same theories has  
led to the concept of extrinsic time
\cite{kuchar2}.  Thus, one may imagine \cite{bar1} a subject ($S$) and reference ($R$) 
GW's which
constructively interfere and give rise to a spacetime holographic image
\cite{bar1,bar2} 
which corresponds to the EM holographs \cite{collier} resulting from the interference of the  $S$ 
and $R$ EM waves  \cite{collier}. This line of reasoning was followed 
in \cite{bar1,bar2} and we
introduce here some results obtained there.  We note that for gravitational
constructive interference the $S$ and $R$ GW's, as for their EM analogues,
should be similar and in phase with each other (see discussion in \cite{bar1,bar2}). 
Thus, the former expressions (\ref{$B_9$})-(\ref{$B_{11}$}) for the GW may be
related not only to either $S$ or $R$ but also, in case of 
 constructive interference, to the combined GW \cite{bar1,bar2} obtained from such an
 interference.  
  The intensity, exposure and transmittance of this combined GW
 were calculated \cite{bar2} in analogy with the corresponding EM quantities. 
 It
 has  also been shown in \cite{bar1} that the holographic images resulting from
 this combined GW correspond to the trapped surfaces
 \cite{eppley,brill2,beig,abrahams} 
 formed from such wave. Thus, since we discuss here these trapped surfaces
 we consider the former expressions (\ref{$B_9$})-(\ref{$B_{11}$})  and the
 following ones as referring to the combined GW obtained from the constructive
 interference of the $S$ and $R$  GW's. We note that 
 it has been shown \cite{tipler,yurtsever} that the
 collision of two
  plane GW's results in a great strengthening (corresponds to constructive
  interference) of the resulting GW  which
    forms, as is the case for strong GW's \cite{mtw,brill2,abrahams,alcubierre}, 
      a 
    singularity \cite{tipler,yurtsever}.\par
 Now, since the  trapped surfaces are embedded in the 
 Euclidean space 
 \cite{mtw,eppley,brill2} 
 one have first to convert \cite{bar1} the
 tensor metric components (see Eqs 
 (\ref{$B_9$})-(\ref{$B_{10}$})) $h^{TT}_{\hat{{\bf n}}_1\hat{{\bf n}}_1}= 
 -h^{TT}_{\hat{{\bf n}}_2\hat{{\bf n}}_2}$, 
 $h^{TT}_{\hat{{\bf n}}_1\hat{{\bf n}}_2}= 
 h^{TT}_{\hat{{\bf n}}_2\hat{{\bf n}}_1}$  
 from  the  
 ${\hat{\bf n}}, \ {\hat{\bf n}}_1, \   {\hat{\bf n}}_2 $ system, into the 
 ${\hat{\bf x}}, \ {\hat{\bf y}}, \  {\hat{\bf z}}$
 Euclidean system.   Thus, substituting \cite{bar1} ${\hat{\bf n}}={\hat{\bf z}},  \ 
 {\hat{\bf n}}_1={\hat{\bf x}}, \ {\hat{\bf n}}_2={\hat{\bf y}}$ 
  one may write 
 the Euclidean metric components as \cite{bar1} 
  \begin{align} &h^{TT}_{{\hat {\bf x}}{\hat{\bf x}}} =\Re 
\biggl( A_+{\bf e}_{+_{{\hat {\bf x}}{\hat{\bf x}}}}e^{-ift}
e^{ik{\bf r}\hat{{\bf z}}} \biggr) = 
 A_+{\bf e}_{+_{{\hat {\bf x}}{\hat{\bf x}}}} \cdot  
\cos(kz -ft) =     \nonumber \\  =& -h^{TT}_{{\hat {\bf y}}{\hat{\bf y}}}=-\Re 
\biggl( A_+{\bf e}_{+_{{\hat {\bf y}}{\hat{\bf y}}}}e^{-ift}
e^{ik{\bf r}\hat{{\bf z}}} \biggr) =  - A_+{\bf e}_{+_{{\hat {\bf y}}{\hat{\bf y}}}} \cdot  
\cos(kz-ft)  \tag{$B_{14}$}  \label{$B_{14}$}   \\  & h^{TT}_{{\hat {\bf x}}{\hat{\bf y}}}=\Re 
\biggl( A_{\times}{\bf e}_{\times_{{\hat {\bf x}}{\hat{\bf y}}}}e^{-ift}
e^{ik{\bf r}\hat{{\bf z}}} \biggr) = 
 A_{\times}{\bf e}_{\times_{{\hat {\bf x}}{\hat{\bf y}}}} \cdot  
\cos(kz- ft) 
=   \nonumber \\  =& h^{TT}_{{\hat {\bf y}}{\hat{\bf x}}}=\Re 
\biggl( A_{\times}{\bf e}_{\times_{{\hat {\bf y}}{\hat{\bf x}}}}e^{-ift}
e^{ik{\bf r}\hat{{\bf z}}} \biggr)  = 
 A_{\times}{\bf e}_{\times_{{\hat {\bf y}}{\hat{\bf x}}}} \cdot  
\cos(kz-ft),  
\nonumber
 \end{align}    
 where ${\bf r}= x{\hat{\bf x}}+y{\hat{\bf y}}+z{\hat{\bf z}}$  and 
  ${\bf e}_{+_{{\hat {\bf x}}{\hat{\bf x}}}}, \  
 {\bf e}_{+_{{\hat {\bf y}}{\hat{\bf y}}}}, \ 
 {\bf e}_{\times_{{\hat {\bf x}}{\hat{\bf y}}}}$ 
 are the Euclidean unit linear-polarization tensors obtained \cite{bar1}  by substituting 
 in 
  Eqs (\ref{$B_7$})-(\ref{$B_8$})  
 ${\hat{\bf n}}={\hat{\bf z}},  \ 
 {\hat{\bf n}}_1={\hat{\bf x}}, \ {\hat{\bf n}}_2={\hat{\bf y}}$.  
 Thus, 
  one may write the
 metrics from  Eq (\ref{$B_2$})  in the TT gauge as \cite{bar1} 
 \begin{align} & (ds^{TT})_{({\hat{\bf x}},{\hat{\bf y}},
{\hat{\bf z}})}^2=h^{TT}_{{\hat {\bf x}}{\hat{\bf x}}}dx^2+
h^{TT}_{{\hat {\bf y}}{\hat{\bf y}}}dy^2+2h^{TT}_{{\hat {\bf x}}{\hat{\bf y}}}dxdy= 
 A_+{\bf e}_{+_{{\hat {\bf x}}{\hat{\bf x}}}} \cdot  
\cos(kz -ft)dx^2+ \nonumber \\  +& A_+{\bf e}_{+_{{\hat {\bf y}{\hat {\bf
y}}}}} \cdot  
\cos(kz-ft)dy^2+2A_{\times}{\bf e}_{\times_{{\hat {\bf x}}{\hat {\bf y}}}} \cdot  
\cos(kz-ft)dxdy =  \tag{$B_{15}$}  \label{$B_{15}$}   \\ 
 =& A_+{\bf e}_{+_{{\hat {\bf x}}{\hat{\bf x}}}} \cdot  
\cos(kz -ft)(dx^2-dy^2)+ 2A_{\times}{\bf e}_{\times_{{\hat{\bf x}{\hat {\bf
y}}}}} \cdot  
\cos(kz-ft)dxdy,   \nonumber \end{align} 
where we use \cite{bar1} 
${\bf e}_{+_{{\hat {\bf x}}{\hat{\bf x}}}}=-{\bf e}_{+_{{\hat {\bf y}}{\hat{\bf
y}}}}$   (see  Eq (\ref{$B_7$})). 
For  calculating the nonfringed embedded surface  
one have to convert \cite{bar1} the last  metric from the 
 $({\hat{\bf x}}, \ {\hat{\bf y}}, \ {\hat {\bf z}})$   system to 
  the cylindrical one
$({\hat{\bf \rho}}, \ {\hat{\bf \phi}}, \ {\hat {\bf z}})$ 
where $x=\rho\cos(\phi), \ y=\rho\sin(\phi), \ z=z$.  Thus, using  the 
trigonometric relations 
$(\cos^2(\phi)-\sin^2(\phi))=\cos(2\phi)$,  
$2\sin(\phi)\cos(\phi)=\sin(2\phi)$   and transforming   the 
 unit polarization tensors 
${\bf e}_{+_{{\hat {\bf x}}{\hat {\bf x}}}}, \  {\bf e}_{+_{{\hat {\bf xy}}}}$ 
to the
corresponding cylindrical ones 
${\bf e}_{+_{{\hat {\bf \rho}}{\hat {\bf \rho}}}}, \ 
{\bf e}_{+_{{\hat {\bf \rho}}{\hat {\bf \phi}}}}$ one obtains \cite{bar1} 
\begin{align}  (&ds^{TT})_{({\hat{\bf \rho}},{\hat{\bf \phi}},
{\hat{\bf z}})}^2=
h^{TT}_{{\hat {\bf \rho}}{\hat{\bf \rho}}}d\rho^2+
h^{TT}_{{\hat {\bf \phi}}{\hat{\bf \phi}}}d\phi^2+
h^{TT}_{{\hat {\bf \rho}}{\hat{\bf \phi}}}d\rho
d\phi= \nonumber \\ 
=& A_+{\bf e}_{+_{{\hat {\bf \rho}}{\hat{\bf \rho}}}} \cdot  
\cos(kz -ft)\bigg(\cos(2\phi)(d\rho^2 -\rho^2d\phi^2)- 
2\rho \sin(2\phi)d\rho d\phi \biggr) 
\tag{$B_{16}$}  \label{$B_{16}$}  \\  +&A_{\times}{\bf e}_{\times_{{\hat {\bf \rho}}{\hat{\bf \phi}}}} \cdot  
\cos(kz-ft)\biggl( \sin(2\phi)(d\rho^2-\rho^2d\phi^2)+
2\rho \cos(2\phi)d\rho d\phi \biggr), \nonumber  \end{align} 
where the unit polarization tensor components  ${\bf e}_{+_{{\hat {\bf \rho}}{\hat {\bf \rho}}}}, \ 
{\bf e}_{+_{{\hat {\bf \rho}}{\hat {\bf \phi}}}}$ are given by \cite{bar1} 
\begin{align} && {\bf e}_{+_{{\hat {\bf \rho}}{\hat {\bf \rho}}}}=
 \cos(2\phi)\biggl( {\bf e}_{\hat {\bf \rho}}\otimes 
{\bf e}_{\hat {\bf \rho}}-{\bf e}_{\hat {\bf \phi}}\otimes 
{\bf e}_{\hat {\bf \phi}}\biggr)-
\sin(2\phi)\biggl( {\bf e}_{\hat {\bf \rho}}\otimes 
{\bf e}_{\hat {\bf \phi}}+{\bf e}_{\hat {\bf \phi}}\otimes 
{\bf e}_{\hat {\bf \rho}}\biggr) \tag{$B_{17}$}  \label{$B_{17}$} \\ 
&& {\bf e}_{\times_{{\hat {\bf \rho}}{\hat {\bf \phi}}}}=
 \sin(2\phi)\biggl( {\bf e}_{\hat {\bf \rho}}\otimes 
{\bf e}_{\hat {\bf \rho}}-{\bf e}_{\hat {\bf \phi}}\otimes 
{\bf e}_{\hat {\bf \phi}}\biggr)+\cos(2\phi)\biggl( {\bf e}_{\hat {\bf \rho}}\otimes 
{\bf e}_{\hat {\bf \phi}}+{\bf e}_{\hat {\bf \phi}}\otimes 
{\bf e}_{\hat {\bf \rho}}\biggr)
 \nonumber \end{align} 
The last tensor components  are obtained by using: (1) $ {\bf e}_{+_{{\hat{\bf x}}{\hat{\bf x}}}}= 
 {\bf e}_{\hat{\bf x}} \otimes {\bf e}_{\hat{\bf x}}- {\bf e}_{\hat{\bf
 y}} \otimes {\bf e}_{\hat{\bf y}}= -{\bf e}_{+_{{\hat{\bf y}}{\hat{\bf
 y}}}}, \ \   
  {\bf e}_{\times_{{\hat{\bf x}}{\hat{\bf y}}}}= 
 {\bf e}_{\hat{\bf x}} \otimes {\bf e}_{\hat{\bf y}}+ {\bf e}_{\hat{\bf
 y}} \otimes {\bf e}_{\hat{\bf x}}= {\bf e}_{\times_{\hat{\bf
 y}{\hat{\bf x}}}}$  which are obtained from  
   Eqs (\ref{$B_7$})-(\ref{$B_8$})  by substituting  \cite{bar1} 
 ${\hat{\bf n}}={\hat{\bf z}},  \ 
 {\hat{\bf n}}_1={\hat{\bf x}}, \ {\hat{\bf n}}_2={\hat{\bf y}}$, (2)
     $ {\bf e}_{\hat {\bf x}}=\cos(\phi){\bf e}_{\hat {\bf \rho}}-\sin(\phi){\bf
  e}_{\hat {\phi}}, \ \ {\bf e}_{\hat {\bf y}}=
  \sin(\phi){\bf e}_{\hat {\bf \rho}}+\cos(\phi){\bf
  e}_{\hat {\phi}}, \  \ {\bf e}_{\hat {\bf z}}= {\bf e}_{\hat {\bf z}} $  
 \cite{spiegel2} and (3) $(\cos^2(\phi)-\sin^2(\phi))=\cos(2\phi)$,  
 $2\cos(\phi)\sin(\phi)=\sin(2\phi)$.   As mentioned after Eq (\ref{e17}),   
 ${\bf e}_{+_{{\hat {\bf \rho}}{\hat {\bf \rho}}}}$, 
 $ {\bf e}_{\times_{{\hat {\bf \rho}}{\hat {\bf \phi}}}}$  as well as 
 ${\bf e}_{\hat {\bf \rho}}\otimes 
{\bf e}_{\hat {\bf \rho}}$, ${\bf e}_{\hat {\bf \rho}}\otimes 
{\bf e}_{\hat {\bf \phi}}$ and ${\bf e}_{\hat {\bf \phi}}\otimes 
{\bf e}_{\hat {\bf \phi}}$ are tensor components in the $\rho\rho$, 
$\rho\phi$   and $\phi\phi$  directions and so, of course,  
they are not tensors proper. This enables one to perform some mathematical
operations on  these components such as comparing them to functions or taking
their  square roots as done, for example,  in Eqs (\ref{e17})-(\ref{e21}). 
 \par 
 Assuming \cite{bar1}, as in \cite{brill1,gentle,eppley},  a non-rotating  
 system  
 so that   $h^{TT}_{{\hat {\bf \rho}}{\hat{\bf \phi}}}$ is identically zero   
one may write    Eq  (\ref{$B_{16}$})  as  \cite{bar1}
 \begin{align}  (&ds^{TT})_{({\hat{\bf \rho}},{\hat{\bf \phi}},
{\hat{\bf z}})}^2=
h^{TT}_{{\hat {\bf \rho}}{\hat{\bf \rho}}}d^2\rho +
h^{TT}_{{\hat {\bf \phi}}{\hat{\bf \phi}}}d^2\phi= 
\tag{$B_{18}$}  \label{$B_{18}$}  \\ 
=& \cos(kz -ft)\biggl(A_+{\bf e}_{+_{{\hat {\bf \rho}}{\hat{\bf \rho}}}}\cos(2\phi)
 +A_{\times}{\bf e}_{\times_{{\hat {\bf \rho}}{\hat{\bf
 \phi}}}}\sin(2\phi)\biggr)\biggl((d\rho^2 - \rho^2d\phi^2)\biggr)= \nonumber 
 \\ = &\cos(kz -ft)\biggl[\frac{\sin(4\phi)}{2}(A_{\times}-A_+)
\left({\bf e}_{\hat {\bf \rho}}\otimes 
{\bf e}_{\hat {\bf \phi}}+{\bf e}_{\hat {\bf \phi}}\otimes 
{\bf e}_{\hat {\bf \rho}}\right)+\nonumber \\  +& 
\left(A_+\cos^2(2\phi)+A_{\times}\sin^2(2\phi)\right)
 \biggl( {\bf e}_{\hat {\bf \rho}}\otimes 
{\bf e}_{\hat {\bf
\rho}}-{\bf e}_{\hat {\bf \phi}}\otimes 
{\bf e}_{\hat {\bf
\phi}}\biggr)\biggr]
 \left(d^2\rho-\rho^2d\phi^2\right) \nonumber 
 \end{align}
 Note that when  $A_{\times}= A_+$   
  the expression $\biggl(A_+{\bf e}_{+_{{\hat {\bf \rho}}{\hat{\bf \rho}}}}\cos(2\phi)
  + A_{\times}{\bf e}_{\times_{{\hat {\bf \rho}}{\hat{\bf
 \phi}}}}\sin(2\phi)\biggr)$ is considerably simplified and reduces \cite{bar1} to 
  $$\biggl(A_+{\bf e}_{+_{{\hat {\bf \rho}}{\hat{\bf \rho}}}}\cos(2\phi)
  + A_{\times}{\bf e}_{\times_{{\hat {\bf \rho}}{\hat{\bf
 \phi}}}}\sin(2\phi)\biggr)=A_+\biggl({\bf e}_{\hat {\bf \rho}}\otimes 
{\bf e}_{\hat {\bf \rho}}-{\bf e}_{\hat {\bf \phi}}\otimes 
{\bf e}_{\hat {\bf \phi}}\biggr)$$ 
    We, now,  find  the relevant  nonfringed embedded  surface and  
assume \cite{bar1}, as done 
 for the nonfringed trapped surfaces resulting from the Brill's  GW  \cite{eppley} (see
 Appendix A here), that its metric is 
 that of a surface of rotation $z(x,y)$ related to Euclidean space  
 as  
  \begin{equation} x^P=f^P(\rho)\cos(\phi), \ \ y^P=f^P(\rho)\sin(\phi), 
  \ \ z^P=h^P(\rho),  
 \tag{$B_{19}$}  \label{$B_{19}$} \end{equation} 
 where the superscript $P$ denotes that we consider plane GW's. 
 Thus,  using 
  Eqs  \ref{$B_{18}$} one may write the metric of the relevant 
  holographic image 
  (trapped surface) as \cite{bar1} (compare with Eq (\ref{$A_{12}$}) in Appendix
  A for the
  Brill case),  
 \begin{align}    
 d&s^2=dx^2+dy^2+dz^2=\biggl((f^P)_{\rho}^2(\rho)+(h^P)_{\rho}^2(\rho)\biggr)d\rho^2+
 (f^P)^2(\rho)d\phi^2=   \tag{$B_{20}$}  \label{$B_{20}$} \\  = &
 h^{TT}_{{\hat {\bf \rho}}{\hat{\bf \rho}}}d^2\rho +
h^{TT}_{{\hat {\bf \phi}}{\hat{\bf \phi}}}d^2\phi  = 
\cos(kz -ft)\biggl(A_+{\bf e}_{+_{{\hat {\bf \rho}}{\hat{\bf \rho}}}}\cos(2\phi)
  + A_{\times}{\bf e}_{\times_{{\hat {\bf \rho}}{\hat{\bf
 \phi}}}}\sin(2\phi)\biggr) \cdot \nonumber \\ \cdot & 
 \biggl(d\rho^2  - 
\rho^2d\phi^2\biggr)= \cos(kz -ft)\biggl[\frac{\sin(4\phi)}{2}(A_{\times}-A_+)
\left({\bf e}_{\hat {\bf \rho}}\otimes 
{\bf e}_{\hat {\bf \phi}}+{\bf e}_{\hat {\bf \phi}}\otimes 
{\bf e}_{\hat {\bf \rho}}\right)+\nonumber \\  + &
\left(A_+\cos^2(2\phi)+A_{\times}\sin^2(2\phi)\right)
 \biggl( {\bf e}_{\hat {\bf \rho}}\otimes 
{\bf e}_{\hat {\bf
\rho}}-{\bf e}_{\hat {\bf \phi}}\otimes 
{\bf e}_{\hat {\bf
\phi}}\biggr)\biggr]
 \left(d^2\rho-\rho^2d\phi^2\right) \nonumber 
 \end{align}
 where Eq (\ref{$B_{17}$}) was used and by  $f^P_{\rho}(\rho), \ h^P_{\rho}(\rho)$ 
 we denote the first 
 derivatives of $f^P(\rho), \ h^P(\rho)$ with respect
 to $\rho$. Using  Eq (\ref{$B_{20}$}) one may determine \cite{bar1} 
 the quantities $f^P(\rho)$, $f^P_{\rho}(\rho)$ and $h^P(\rho)$ which
 defines the intrinsic geometry of the nonfringed  trapped surface 
 (gravitational holograph) 
 \begin{align} & f^P(\rho)= \rho\biggl[\cos(kz
 -ft)\biggl\{\frac{\sin(4\phi)}{2}(A_+-A_{\times})
\left({\bf e}_{\hat {\bf \rho}}\otimes 
{\bf e}_{\hat {\bf \phi}}+{\bf e}_{\hat {\bf \phi}}\otimes 
{\bf e}_{\hat {\bf \rho}}\right)+\nonumber \\  +& 
\left(A_+\cos^2(2\phi)+A_{\times}\sin^2(2\phi)\right)
 \biggl( {\bf e}_{\hat {\bf \phi}}\otimes 
{\bf e}_{\hat {\bf
\phi}}-{\bf e}_{\hat {\bf \rho}}\otimes 
{\bf e}_{\hat {\bf
\rho}}\biggr)\biggr\}\biggr]^{\frac{1}{2}} \nonumber  \\ & 
 f^P_{\rho}(\rho)= \biggl[\cos(kz
 -ft)\biggl\{\frac{\sin(4\phi)}{2}(A_+-A_{\times})
\left({\bf e}_{\hat {\bf \rho}}\otimes 
{\bf e}_{\hat {\bf \phi}}+{\bf e}_{\hat {\bf \phi}}\otimes 
{\bf e}_{\hat {\bf \rho}}\right)+\nonumber \\  +& 
\left(A_+\cos^2(2\phi)+A_{\times}\sin^2(2\phi)\right)
 \biggl( {\bf e}_{\hat {\bf \phi}}\otimes 
{\bf e}_{\hat {\bf
\phi}}-{\bf e}_{\hat {\bf \rho}}\otimes 
{\bf e}_{\hat {\bf
\rho}}\biggr)\biggl\}\biggr]^{\frac{1}{2}}   \tag{$B_{21}$}  \label{$B_{21}$} \\ 
& h^P(\rho)=\int d\rho\biggl[\cos(kz -ft)\biggl\{\frac{\sin(4\phi)}{2}(A_{\times}-A_+)
\left({\bf e}_{\hat {\bf \rho}}\otimes 
{\bf e}_{\hat {\bf \phi}}+{\bf e}_{\hat {\bf \phi}}\otimes 
{\bf e}_{\hat {\bf \rho}}\right)+\nonumber \\  +& 
\left(A_+\cos^2(2\phi)+A_{\times}\sin^2(2\phi)\right)
 \biggl( {\bf e}_{\hat {\bf \rho}}\otimes 
{\bf e}_{\hat {\bf
\rho}}-{\bf e}_{\hat {\bf \phi}}\otimes 
{\bf e}_{\hat {\bf
\phi}}\biggr)\biggr\}-(f^P)_{\rho}^2(\rho)\biggr]^{\frac{1}{2}}
\nonumber  \end{align}

 \markright{REFERENCES}
 
  \bigskip \bibliographystyle{plain}

\begin{thebibliography}{99}
\bibitem{camera} The kind of digital camera used was HP Photosmart M22 and the
photographs taken were in JPEG format which  were converted to the PS one.
\end{thebibliography}

\end{document}